\documentclass[10pt,sigconf]{acmart}

\usepackage{xspace}
\usepackage{url}
\usepackage{bm}
\usepackage{epsfig,bm,graphicx}
\usepackage{booktabs}
\usepackage{balance}  
\usepackage{xcolor}
\usepackage{amsmath}
\usepackage{makecell}

\newcommand{\stitle}[1]{\vspace{1ex}\noindent{\bf #1}}

\usepackage[linesnumbered,ruled]{algorithm2e}
\usepackage{setspace}
\SetKwRepeat{Do}{do}{while}
\usepackage{cleveref}
\usepackage{enumitem}
\usepackage{textcomp}
\usepackage{multirow}
\usepackage{bbm}
\usepackage{adjustbox}
\usepackage[titletoc]{appendix}
\usepackage[thinc]{esdiff}

\newcommand{\revisenew}[1]{#1}

\newcommand{\system}{{\sc ZeroER}\xspace}

\newcommand*{\rowstyle}[1]{
  \gdef\@rowstyle{#1}%
  \@rowstyle\ignorespaces%
}

\newcolumntype{=}{
  >{\gdef\@rowstyle{}}%
}

\newcolumntype{+}{
  >{\@rowstyle}%
}

\copyrightyear{2020} 
\acmYear{2020} 
\setcopyright{acmcopyright}
\acmConference[SIGMOD'20]{Proceedings of the 2020ACM SIGMOD International Conference on Management of Data}{June 14--19,2020}{Portland, OR, USA}
\acmBooktitle{Proceedings of the 2020 ACM SIGMOD International Conference onManagement of Data (SIGMOD'20), June 14--19, 2020, Portland, OR, USA}
\acmPrice{15.00}
\acmDOI{10.1145/3318464.3389743}
\acmISBN{978-1-4503-6735-6/20/06}

\begin{document}

\title{ZeroER: Entity Resolution using Zero Labeled Examples}

\author{Renzhi Wu}
\orcid{0002-9144-8999}
\affiliation{%
  \institution{Georgia Institute of Technology}
}
\email{renzhiwu@gatech.edu}

\author{Sanya Chaba}
\affiliation{%
  \institution{Georgia Institute of Technology}
}
\email{sanyachaba@gatech.edu}

\author{Saurabh Sawlani}
\affiliation{%
  \institution{Georgia Institute of Technology}
}
\email{sawlani@gatech.edu}

\author{Xu Chu}
\affiliation{%
  \institution{Georgia Institute of Technology}
}
\email{xu.chu@cc.gatech.edu}

\author{Saravanan Thirumuruganathan}
\affiliation{%
  \institution{QCRI, HBKU}
}
\email{sthirumuruganathan@hbku.edu.qa}

\renewcommand{\shortauthors}{Wu, et al.}

\begin{abstract}

Entity resolution (ER) refers to the problem of matching records in one or more relations that refer to the same real-world entity.
While supervised machine learning (ML) approaches achieve the state-of-the-art results,
they require a large amount of labeled examples that are expensive to obtain and often times infeasible.
We investigate an important problem that vexes practitioners:
is it possible to design an effective algorithm for ER that requires \emph{Zero} labeled examples,
yet can achieve performance comparable to supervised approaches?
In this paper, we answer in the affirmative through our proposed approach dubbed \system.

Our approach is based on a simple observation ---
\emph{the similarity vectors for matches should look different from that of unmatches}. Operationalizing this insight requires a number of technical innovations.
First, we propose a simple yet powerful generative model based on Gaussian Mixture Models
for learning the match and unmatch distributions.
Second, we propose an adaptive regularization technique customized for ER
that ameliorates the issue of feature overfitting. 
Finally, we incorporate the transitivity property into the generative model in a novel way
resulting in improved accuracy. On five benchmark ER datasets, we show that  \system  greatly outperforms existing unsupervised approaches and achieves comparable performance to supervised approaches.


\end{abstract}



\maketitle
\vspace{0mm}
\section{Introduction}
\label{sec:intro}
Entity resolution (ER) -- also known as duplicate detection or record matching -- refers to the problem of identifying tuples in one or more relations that refer to the same real world entity.
For example, an e-commerce website would want to identify duplicate products (such as from different suppliers) so that they could all be listed in the same product page.
ER has been extensively studied in many research communities, including databases, statistics, NLP, and data mining (e.g., see multiple surveys~\cite{DBLP:conf/sigmod/KoudasSS06, DBLP:journals/tkde/ElmagarmidIV07, DBLP:journals/pvldb/DongN09, DBLP:series/synthesis/2010Naumann, DBLP:journals/pvldb/GetoorM12,herzog2007data}).

\begin{figure*}[t]
  \centering
  \centerline{\epsfig{figure=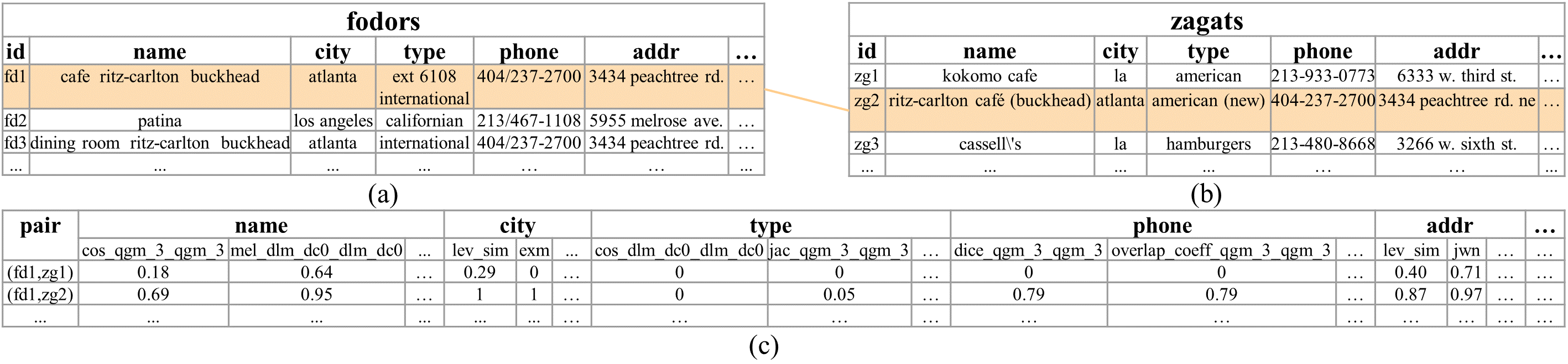, width=18cm}}
  \vspace{-4mm}
  \caption{ER on a benchmark restaurant dataset~\cite{uci} with two tables (a) fodors table and (b) zagats table. (c) shows similarity vectors (aka feature vectors) generated by the open-source ER package Magellan~\protect\cite{konda2016magellan}.}
  \label{fig:ER_illu}
\vspace{-2mm}
\end{figure*}

The standard approach for entity resolution is to use textual similarities between two tuples to determine if they are a match or unmatch~\cite{DBLP:journals/tkde/ElmagarmidIV07}.
Informally, in \Cref{fig:ER_illu}(a)(b), the matching pair (fd1, zg2) is textually similar, while the non-matching pair (fd1, zg1) is not.
Typically,  a vector of similarity scores (aka, feature vector) is computed for a tuple pair by applying various similarity functions (such as edit distance, or jaccard similarity) on corresponding attributes. Multiple similarity scores can be used for a single attribute.
\Cref{fig:ER_illu}(c) shows a subset of the similarity scores computed by the open-source ER package Magellan\cite{konda2016magellan}. For example, the feature cos\_qgm\_3\_qgm\_3 represents the cosine similarity between 3-grams on the name attribute between a pair of tuples.
Given the computed feature vectors, ER is typically formulated as a binary classification problem where a classifier is trained to determine whether a tuple pair is a match or an unmatch based on its similarity vector.

\stitle{The Need for \system.}
Supervised machine Learning (ML) approaches provide the state-of-the-art results for ER~\cite{deeper,anhaisigmod2018,kopcke2010evaluation,dong2018data}.
Powerful techniques such as deep learning have a voracious appetite for large amounts of labeled training data.
For example, both DeepER~\cite{deeper} and DeepMatcher~\cite{anhaisigmod2018}
require hundreds to thousands of labeled examples.
Even non-deep learning based methods require hundreds of labeled examples~\cite{kopcke2010evaluation}.
It has been reported~\cite{dong2018data} that achieving F-measures of $\sim$99\% with random forests can
require up to 1.5M labels even for relatively clean datasets.
Hence, generating large amounts of training data is a time consuming and challenging task even for domain experts.

Furthermore, organisations often have many relations to be de-duplicated (e.g., General Electric has 75 different procurement systems~\cite{stonebraker2018data}), and generating labeled data becomes a huge issue.
Due to this, there has been intense interest from the database community in
self-service data preparation that  requires limited efforts from domain scientists.
Motivated by these considerations,  we develop a \system approach that can perform ER with comparable performance to supervised approaches, but with zero labeled examples.


\vspace{0mm}
\stitle{Key Insights and Challenges.} Designing a \system approach that achieves competitive performance across different datasets is extremely challenging. We highlight three fundamental ideas that make this possible.


\noindent \underline{(1) Generative Modelling.}
Supervised approaches learn to tell apart matches and unmatches from the large amount of labeled training data.
However, \system does not have this useful information and
has to design a mechanism to distinguish the matches from unmatches.
\system leverages a blindingly simple yet powerful observation:
\emph{the similarity vectors (aka feature vectors) for matches should look different from the similarity vectors for unmatches.}

\begin{figure}[htb]
\vspace{-3mm}
\begin{minipage}[b]{1\linewidth}
  \centering
  \centerline{\epsfig{figure=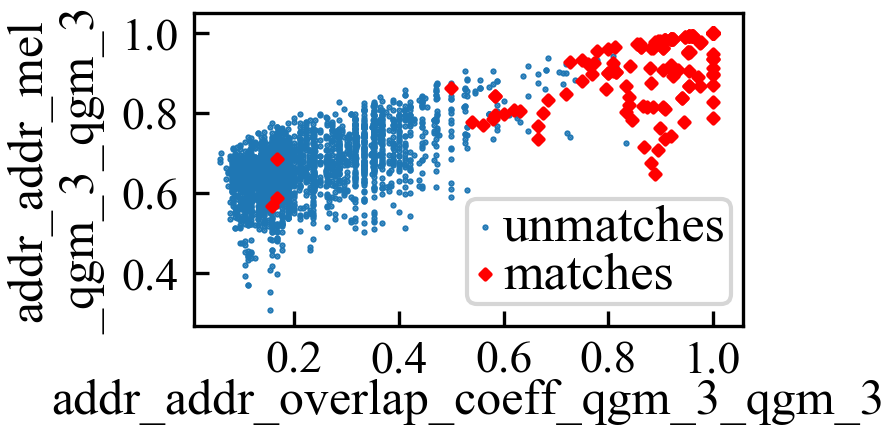,height=2.5cm}}
  \vspace{-4mm}
  \caption{Distribution of matches and unmatches on two features in the fodors-zagats dataset.}
  \label{fig:matches_similar}
\end{minipage}
\vspace{-10mm}
\end{figure}

We plot in \Cref{fig:matches_similar} the distribution of feature values for the fodors-zagats dataset, using ground-truth matches and unmatches. For ease of visualization, we only show the distribution along two feature dimensions. As we can see, the similarity values of these two features for the matches concentrate in the upper right corner, while the values for unmatches in the bottom left corner.

This observation inspires us to use \textit{generative modelling for the ER problem} --- the feature vector $\textbf{x}$ of a match can be assumed to be generated according to one distribution (termed the M-Distribution), and the feature vector $\textbf{x}$ of an unmatch can be assumed to be generated according to a different distribution (termed the U-Distribution). 


If we can learn the probability density functions (PDFs) of two distributions, i.e. $p(\textbf{x} | y = M)$ and $p(\textbf{x} | y = U)$, and the prior probability of a tuple pair being a match $\pi_M = p(y = M)$, or a non-match $\pi_U = p (y = U) = 1 - \pi_M$, then determining whether a given tuple pair should be classified as match is equivalent to computing the posterior probability $p(y = M | \textbf{x})$ for its feature vector $\textbf{x}$ using the Bayes rule:
\begin{small}
\vspace{-1mm}
\begin{align}
    \label{eq:bayesRulePosteriorGeneral}
  p(y = M|\textbf{x}) &= \frac{p(\textbf{x},y = M)}{p(\textbf{x})} = \frac{p(\textbf{x},y = M)}{p(\textbf{x}, y = M) + p(\textbf{x}, y = U)} \nonumber \\
     &= \frac{\pi_M \times p(\textbf{x}|y = M)}{\pi_M \times p(\textbf{x}|y = M) + \pi_U \times p(\textbf{x}|y = U)}
\end{align}
\vspace{-3mm}
\end{small}

Accurately learning the prior $\pi_M$, the M-Distribution $p( \textbf{x} | y = M )$, and the U-Distribution $p( \textbf{x} | y = U )$ -- with Zero labeled data -- is the key to achieve high-quality predictions.


Let $\theta_M$ (resp. $\theta_U$) denote the parameters of the M-Distribution (resp. U-Distribution). Different choices of $\theta_M$ and $\theta_U$ correspond to different generative models. We choose to use Gaussian distributions for both the M- and the U-Distribution, as the Gaussian distribution is known to be the default distribution used in various areas to represent real-valued random variables with unknown distributions~\cite{lyon2013normal}. In our case, the two distributions would be multi-dimensional Gaussian distributions parameterized by a mean vector and a co-variance matrix $\theta_C = \{\mu_C, \Sigma_C\}$, where $C \in \{M, U\}$.

This reduces the generative model to the Gaussian Mixture Model (GMM) with two components, and all parameters $\{ \pi_M, \mu_C, \Sigma_C\},  C \in \{M, U\}$ can then be learned by maximizing the data likelihood via an expectation-maximization algorithm without any labeled data~\cite{dempster1977maximum}. However, using the naive GMM as the generative model for ER produces inferior results, since it fails to recognize one important characteristic of the ER problem, namely, the number of matches is usually a tiny fraction of the total number of tuple pairs. The lack of matching pairs makes learning $\theta_M$ particularly difficult.

\begin{example}
\vspace{-2mm}
In the fodors-zagats dataset, we have 112 matches in the ground truth, and we have seven attributes. An out-of-box invocation of the open-source Magellan package~\protect\cite{konda2016magellan} generated 68 features for those seven attributes, which means we have 2346 ($=68 + 68*67/2$) parameters in $\Sigma_M$. Using a small number of examples (e.g., 112) to estimate a large co-variance matrix (e.g., 2346 parameters) causes inaccurate estimates~\cite{Dempster1972Mar,Velasco}.



\vspace{-2mm}
\end{example}{}

In Section~\ref{sec:autoerModel}, we show how existing approaches to reduce the number of parameters in the naive GMM is not suitable for ER, and we will introduce \system's generative model that includes two novel and ER-specific adaptations to the naive GMM. Our generative model would only require $(4d + 1)$ total number of parameters for a dataset with $d$-dimensional features.





\noindent \underline{(2) Feature Regularization.} In supervised ML settings, overfitting is a well-known problem that causes an ML model to  fit too closely or exactly to a particular set of data (usually the training set), and may therefore fail to fit additional data or predict future observations reliably well (usually the test set). As we were developing and testing generative models for \system, we discovered a similar ``overfitting'' problem that causes the generative model to fit too closely or exactly to some particular features, such that other features play almost no role in making predictions.


\begin{example}
\label{ex:overfitting}

For simplicity, let us illustrate the overfitting problem with two features in the fodors-zagats dataset. The first feature $f_1: city\_city\_exm$ represents whether the city attribute of a tuple pair is the same or not, and the second feature $f_2: phone\_phone\_jar$ represents the jaccard similarity between the phone attribute of a tuple pair. Figure~\ref{fig:overfitting} shows the frequency of tuple pairs taking different values of these two features using ground-truth matches and unmatches. As we can see, matches and unmatches may have the same or different city attribute; and matches usually have a much higher jaccard similarity on phone attribute than unmatches.

\begin{figure}[htb]
  \vspace{-4mm}
\begin{minipage}[b]{1.0\linewidth}
  \centering
  \centerline{\epsfig{figure=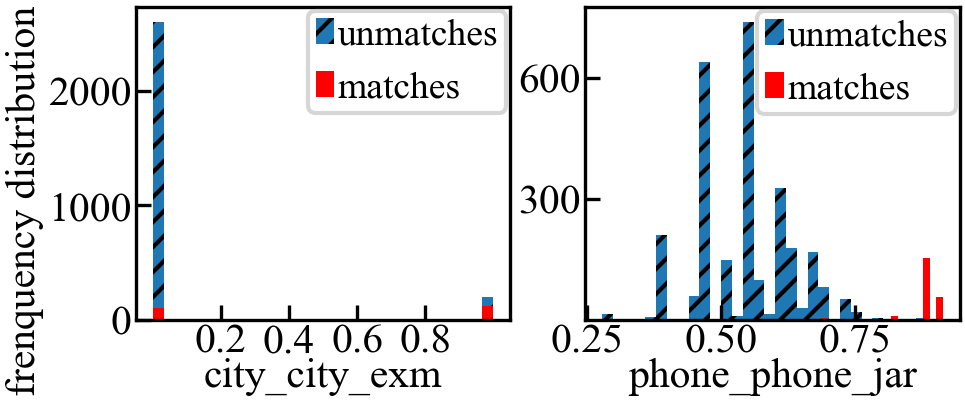,height=2.5cm}}
  \vspace{-4mm}
  \caption{Distribution of two features in the fodors zagats dataset.}
  \label{fig:overfitting}
\end{minipage}
\vspace{-10mm}
\end{figure}

For ease of illustration, let us assume that $f_1$ and $f_2$ are independent (though the same overfitting occurs even if features are dependent). Therefore, the M-Distribution (similarly, the U-Distribution) can be written as $p(\textbf{x} | y = M) = p(\textbf{x}[f_1]| y = M) \times p(\textbf{x}[f_2]| y = M)$, i.e., we can fit the two features independently.

Fitting a GMM \textit{without labeled data} on $f_1$ would give the following two components:  the first component has a mean of 1.0 and a variance of 0, and the second component has a mean of 0.0 and a variance of 0, as $f_1$ has two clearly separated density regions. As we know matches generally have higher similarities than unmatches, we conclude that  the first component corresponds to  $p(\textbf{x}[f_1]| y = M)$, and the second component corresponds to $p(\textbf{x}[f_1]| y = U)$. Similarly, suppose fitting a GMM \textit{without labeled data} on $f_2$ gives the following: $p(\textbf{x}[f_2]| y = M)$ has a mean of 0.9 and a variance of 0.1. $p(\textbf{x}[f_2]| y = U)$ has a mean of 0.6 and a variance of 0.2.


If we use the above learned distributions to make predictions, a feature vector $\textbf{x}$ with $x[f_1] = 1.0$ will be surely predicted as a match, regardless of the value of $x[f_2]$. This is because $p(\textbf{x}[f_1]| y = M)$ will have an infinite density when  $x[f_1] = 1.0$ due to its 0 variance. Therefore, $p(\textbf{x}| y = M)$ will be infinity, which gives $p(y = M | \textbf{x}) = 1.0$ using \Cref{eq:bayesRulePosteriorGeneral}. Similarly, a feature vector $\textbf{x}$ with $x[f_1] = 0$ will be surely predicted as a unmatch, regardless of the value of $x[f_2]$.

The GMM will predict matches or unmatches solely based on $f_1$, and $f_2$ plays no role. This is problematic, as we know in the ground truth (see Figure~\ref{fig:overfitting}), there are matches with $x[f_1] = 0$ and unmatches with  $x[f_1] = 1$ . If the model took into account $f_2$ properly, then we may be able to prevent these erroneous predictions.






\vspace{-3mm}
\end{example}{}

While the above example shows an extreme case where one feature entirely dictates the prediction, in practice, we see many features with extremely small variance that essentially dominate the computation, leaving other features to have little impact on the prediction. \revisenew{Note that this feature overfitting problem is caused by degenerate features, so it happens regardless of the amount of data.}

Our observed feature overfitting in the extreme case, where one feature entirely dictates the prediction, is similar  to the known \textit{singularity} problem in pattern recognition where the determinant of a covariance matrix is zero~\cite{bishop2006pattern}. A common solution to address the singularity problem is to add a small constant value to the diagonal entries of the covariance matrix~\cite{gmm_sklearn}. While this avoids having infinite values in the computation, it does not address our overfitting problem and certain features may still dominate the prediction.  

In Section~\ref{sec:feature_regularization}, we propose an ER-specific adaptive regularization strategy that regularizes each feature differently, depending on the overlap between the M- and U-Distribution in each feature. Our proposed regularization strategy ensures that every feature can contribute to making predictions.



\noindent \underline{(3) Incorporating Transitivity.}
ER exhibits the \textit{transitivity} property:
if both $(t_1, t_2)$ and $(t_1, t_3)$ are matches, then $(t_2, t_3)$ is also a match.
Intuitively, incorporating transitivity into the generative model should increase its performance, since it provides additional information.


\begin{example}
\vspace{-2mm}
In \Cref{fig:ER_illu}, both (fd1,zg2) and (fd3, zg2) have high similarity, so a model tends to classify both as matches. In the ground truth, (fd1,zg2) is indeed a true match, while (fd3, zg2) is a false positive. We can see that fd1 and zg2 are the same cafe in a hotel, while fd3 is the dining room in that hotel. Without such domain knowledge, it is very difficult for a model to recognize that (fd3, zg2) is not a match.

However, this false positive can be detected using transitivity. Classifying (fd1,zg2) and (fd3, zg2) as matches will make (fd1, fd3) a match too through transitivity. If we know the fodors table is duplicate free (as we show later, we can leverage transitivity without assuming either input table to be duplicate free), then (fd1, fd3) cannot be a match. This means that one of (fd1,zg2) and (fd3, zg2) has to be a false positive. Since (fd1,zg2) is more similar than (fd3, zg2), it will be easy for a model to conclude that (fd3, zg2) is not a match.
\vspace{-5mm}
\end{example}{}

The supervised approaches make predictions on each unlabeled tuple pair independently and can only address transitivity violations as a postprocessing step.
%
In contrast, we incorporate transitivity in the model learning process, which gives a significant boost in model performance. Intuitively, transitivity serves as  constraints on the labeling possibilities of all tuple pairs. Solving the ``constrained optimization'' problem is highly non-trivial as it is a non-convex problem with quadratic number of constraints with respect to the number of tuple pairs.

In Section~\ref{sec:transitivity}, we introduce a novel way of incorporating transitivity in the generative model, and propose optimizations to solve it with minimal additional computation cost.







\stitle{Contributions.}
We propose \system for entity resolution using zero labeled examples with the following features:
\begin{itemize}[leftmargin=0mm]
\vspace{-1mm}
  \item[] \underline{(1) Unsupervised}: \system requires minimal human input. Users only need to provide two input tables $L$ and $R$, similarity measures used to create the feature matrix, and optionally a blocking function --- \system is then able to determine matches and non-matches automatically with zero labeled training data.

  \item[] \underline{(2) High Quality}: As \system includes several innovations that exploit ER-specific properties, it produces high-quality results across several distinct domains. Specifically, we show that \system i)  greatly outperforms  unsupervised baseline methods; and ii) achieves comparable results against supervised and active learning methods, but saves substantial human labeling efforts.

  \item[] \underline{(3) Efficient}:
  We design an efficient iterative expectation-maximization based algorithm
  for learning the parameters of M- and U-distributions.
  We ensure that materializing our insights, including
  feature regularization and incorporating transitivity, do not come at the expense of computational performance.




\vspace{-1mm}
\end{itemize}

\vspace{-1mm}
\section{Preliminaries}
\label{sec:preliminaries}

We formally define entity resolution in Section~\ref{subsec:entityResolution} and describe the generative modelling idea for ER and how to train a generative model in Section~\ref{sec:generative_model}.
\vspace{-2mm}
\subsection{Entity Resolution Problem Definition}
\label{subsec:entityResolution}
An entity is a distinct real-world object such as a customer, an organization, a publication etc.
We are given two relations $T$ and $T^{\prime}$ with attributes $\{A_1, A_2, \ldots, A_m\}$.
The entity resolution (ER) problem~\cite{DBLP:journals/tkde/ElmagarmidIV07} seeks to identify all pairs of tuples $(t, t^{\prime})$,
where $t \in T, t^{\prime} \in T^{\prime}$, that refer to same real-world entity.
A pair of tuples is said to be a \emph{match} (denoted as M) (resp.  \emph{unmatch} (denoted as U) )
when they refer to the same (resp.  different) real-world entity. When $T = T'$, this is also known as \textit{data deduplication}; and when $T \neq T'$, this is also known as \textit{record linkage}.
Figure~\ref{fig:ER_illu} provides an illustration where we wish to identify which pairs of records from Fodors and Zagat refer to the same restaurants.

To reduce  the quadratic number of tuple pairs an ER problem has ($|T| \times |T'| $), a \textit{blocking function} ${\mathcal B}$  typically is used to filter out those tuple pairs that are highly unlikely to be matches (e.g., tuple pairs with different values in the state attribute will be not be compared), and obtain a candidate set $C_s \subseteq T \times T^{\prime}$ that is usually much smaller. Designing an effective blocking strategy is an orthogonal research problem, and many prior work has been dedicated to this~\cite{whang2009entity,papadakis2013meta,papadakis2016comparative}. The blocking function ${\mathcal B}$ is an optional input to our system, and $C_s = T \times T'$ if no blocking is provided.






To determine the M or U label for every tuple pair, one needs to generate  similarity vectors (aka feature vectors) for every tuple pair in $C_s$ by applying a set of similarity functions ${\mathcal F}$. Given an ER task, typically, ${\mathcal F}$ is either provided by domain experts, or generated according to some default feature engineering strategy implemented in ER packages. \system can work with any ${\mathcal F}$. \Cref{fig:ER_illu}(c) shows a sample of features generated by the open-source Magellan package\cite{konda2016magellan}.

\stitle{Formal Problem Definition.} Given two relations $T$ and $T^{\prime}$,  a set of similarity functions ${\mathcal F}$ for this dataset, and an optional blocking function ${\mathcal B}$, the goal of \system is to assign a binary label $y \in \{M,U\}$ for every pair in $T \times T'$ without any labeled data.

We apply ${\mathcal B}$ (if available) to obtain a set of $N$ tuple pairs $C_s \subseteq T \times T'$, and then apply ${\mathcal F}$ to obtain a feature vector $\textbf{x} = (x^1, x^2, \ldots, x^d)$ for every tuple pair in $C_s$. We use $y$ to denote the unknown label for every feature vector $\textbf{x}$.




\vspace{-2mm}
\subsection{Generative Modelling for ER}
\label{sec:generative_model}



As discussed, we propose to solve the ER problem using generative modelling in the absence of labeled examples. Specifically, we consider the features and label $(\textbf{x},y)$ of a tuple pair  to be generated as follows:
\begin{enumerate}[leftmargin=*]
    \itemsep0em
    \item Generate the value for $y$ (M or U) based on the Bernoulli distribution parameterized by $\pi_M$.
        Intuitively, the process tosses a coin that comes heads with probability $\pi_M$.
        If it comes heads, it chooses $y = M$, else $y = U$.
    \item Generate the feature vector $\bf{x}$ from the selected distribution.
        For example, if $y = M$, then $\textbf{x}$ is sampled according to the M-Distribution  $p(\textbf{x}| y = M)$, parameterized by $\theta_M$; else $\textbf{x}$ is sampled according to the U-Distribution  $p(\textbf{x}| y = U)$, parameterized by $\theta_U$.
\end{enumerate}

The generative model is governed by its set of parameters $\Theta = \{\theta_M, \theta_U, \pi_M\}$. If we can learn $\Theta$, then  we can classify any tuple pair based on its similarity vector $\textbf{x}$.
\revisenew{Specifically, the classification problem is to determine whether $p(y=M|\textbf{x}) > 0.5$. In supervised learning, a functional mapping between the feature space and label space is learned using labeled examples, which directly gives $p(y = M | \textbf{x})$. 
In \system, instead of learning $p(y = M | \textbf{x})$ directly, we aim to learn the prior probability $p(y = M)$,  the M-Distribution $p(\textbf{x} | y = M)$, and the U-Distribution $p(\textbf{x} | y = U)$, which we can use to compute $p(y = M | \textbf{x})$ via ~\Cref{eq:bayesRulePosteriorGeneral}.
}

\stitle{Learning Parameters of Generative Model.}
Given $N=|C_s|$ tuple pairs $(\textbf{x}_i,y_i)$, $i \in [1,N]$, where $\textbf{x}_i$ is observed and $y_i$ is unknown, a common way to estimate $\Theta$ is by maximizing the \textit{log data likelihood function} (the log is taken for computational convenience):
\vspace{-1mm}
\begin{small}
\begin{align}
\label{eq:complete_logdatalikelihood}
    L(\Theta) =& \log \Pi_{i=1}^{N} p(\textbf{x}_i, y_i|\Theta)  = \Sigma_{i=1}^N \log \big( p(y_i|\Theta)p(\textbf{x}_i|\Theta, y_i) \big)\nonumber \\
     =& \Sigma_{i=1}^N \Sigma_{C \in \{M,U\}} \mathbbm{1}_{y_i = C}\log \big( \pi_C p(\textbf{x}_i|\theta_C) \big)
\end{align}
\end{small}
where $\mathbbm{1}_{.}$ is the identity function that evaluates to 1 if the condition is true and 0 otherwise; and $p(\textbf{x}_i|\theta_C)$ is short for $p(\textbf{x}_i|y_i=C)$ parameterized by $\theta_C$.

Since $y_i$ is unknown,~\Cref{eq:complete_logdatalikelihood} cannot be optimized directly. The EM algorithm is the canonical algorithm to use in the case of unobserved variables~\cite{dempster1977maximum}.
Each iteration of the EM algorithm consists of two steps: an Expectation (E)-step and a Maximization (M)-step.
Intuitively, the E-step determines what is the (soft) class assignment $y_i$ for every tuple pair based on the parameter estimates from last iteration $\Theta^{t-1}$. In other words, E-step computes the posterior probability $\gamma_{i,M} = p(y_i = M | \textbf{x}_i, \Theta^{t-1})$. The M-step takes the new soft class assignments and re-estimates all parameters $\Theta^{t}$.
Formally, the two steps are as follows:

\begin{enumerate}[leftmargin=*]
  \item \textbf{E Step.} Given the parameter estimates from the previous iteration $\Theta^{t-1}$, compute the posterior probabilities:
\begin{small}
\vspace{-2mm}
\begin{align}
	\label{eq:e_step}
    \gamma_{i,M}^t = & p(y_i = M | \textbf{x}_i, \Theta^{t-1})\\ \nonumber
      =& \frac{\pi_M^{t-1} \times p(\textbf{x}_i|\theta_M^{t-1})}{\pi_M^{t-1} \times p(\textbf{x}_i|\theta_M^{t-1}) + (1-\pi_M^{t-1}) \times p(\textbf{x}_i|\theta_U^{t-1})}
\end{align}
\vspace{-2mm}
\end{small}
  \item \textbf{M Step.}Given the new soft class assignments as defined by $\gamma_{i,M}^t$, re-estimate $\Theta$ by maximizing the following expected log likelihood function:
  \begin{small}
    \begin{align}
    \label{eq:m_step}
  \mathbb{E}\{L(\Theta) \} =& \Sigma_{i=1}^N \Sigma_{C \in \{M,U\}} \gamma_{i,C}^t\log \big( \pi_C p(\textbf{x}_i|\theta_C) \big)
  \end{align}
  \end{small}
In other words, $\Theta^t = \text{arg} \max_{\Theta}\mathbb{E}\{L(\Theta)\}$.
\end{enumerate}


\vspace{-1mm}
\stitle{GMM for ER.}
%
The generative model becomes the Gaussian Mixture Model (GMM) with two mixture components when choose the M and U distribution ($p(\textbf{x}| y = M)$ and $p(\textbf{x}| y = U)$) to be Gaussian.
Gaussian distribution often serves as an effective default distribution for numerical random variables whose distributions are not known~\cite{lyon2013normal}.
Furthermore, there exist closed-form solutions for estimating the parameters of a Gaussian distribution given a set of data points.
Under GMM, the parameters for M-distribution and U-distribution are $\theta_C = \{\mu_C, \Sigma_C\}$,
where $C \in \{M, U\}$, $\mu_C$ is the mean vector, and $\Sigma_C$ is the covariance matrix.

The EM algorithm can now be instantiated with the Gaussian PDF $p(\textbf{x}_i | \theta_C) = \frac{1}{(2\pi)^{d/2}} \frac{1}{\text{det}(\Sigma_C)^{1/2}} \exp\{- \frac{1}{2} (\textbf{x}_i - \mu_C)^T \Sigma_C^{-1} (\textbf{x}_i-\mu_C)\}$. The E-step can be directly calculated. The M-step has a closed-form solution~\cite{bishop2006pattern} as follows:
\begin{small}
\begin{align}
\label{eq:optimal_sigma_mu}
	N_C =& \Sigma_{i=1}^{N} \gamma_{i,C};\  \pi_C =  N_C/N;\ \mu_C= \Bar{\textbf{x}}_C=1/N_C\Sigma_{i=1}^N \gamma_{i,C}\textbf{x}_i\\ \nonumber
	  \Sigma_C =&S_C= 1/N_C\Sigma_{i=1}^N\gamma_{i,C} (\textbf{x}_i-\Bar{\textbf{x}}_C) (\textbf{x}_i-\Bar{\textbf{x}}_C)^T
\end{align}
\end{small}
where $C \in \{M, U\}$. In other words, $\mu_C$ and $\Sigma_C$ are assigned to be the weighted sample mean $\Bar{\textbf{x}}_C$ and weighted sample covariance $S_C$ respectively in the M-step.

\section{The ZeroER Generative Model}
\label{sec:autoerModel}

GMM as the generative model for ER described in Section~\ref{sec:preliminaries} is a promising start.
However, as we shall show, blindly applying it for ER produces sub-optimal results.
The key issue is that GMM is not adapted to handle properties unique to ER, which we address next.

\vspace{-2mm}
\subsection{Data Deficiency}
In a typical ER dataset, the number of matches is dramatically less than the number of unmatches by many orders of magnitude.
This impacts the accuracy of parameter estimation for the generative model.
Specifically, the number of parameters to be learned for the covariance matrix $\Sigma_M$ is ${d \choose 2}$ under GMM.
Recall from Section~\ref{sec:intro} that for many datasets such as Fodors-Zagat,
the value of ${d \choose 2}$ can even be higher than the number of matches, which would lead to an inferior estimate of $\Sigma_M$. Specifically, this would cause a large systematic distortion on the eigenstructure of $\Sigma_M$~\cite{Dempster1972Mar,Velasco}, and hence affecting the accuracy of the M-Distribution. 

\stitle{The existing approaches.}
Existing approaches for tackling data deficiency makes simplifying assumptions
to reduce the number of parameters. As provided in the well-known sklearn package~\cite{gmm_sklearn}, the two most common approaches are (1) assuming feature independence,i.e. $\Sigma_M$( and also $\Sigma_U$) is diagonal (non-diagonal entries are 0). (2) assuming the covariance matrix can be shared across all components, i.e. $\Sigma_M = \Sigma_U$.
While these two approaches reduce the number of parameters significantly, as we shall show next
and also in experiments, they are unsuitable for ER.
We propose two ER-specific ways to reduce the number of parameters next.  

\begin{figure}[htb]
\vspace{-3mm}
\begin{minipage}[b]{1.0\linewidth}
  \centering
  \centerline{\epsfig{figure=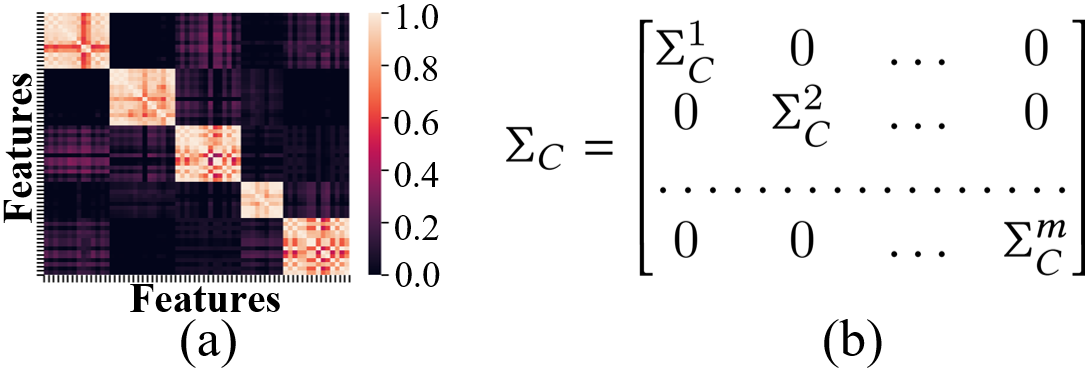, width=6cm}}
  \vspace{-4mm}
  \caption{(a) Heat map of correlation between features using ground-truth matches in the fozors-zagats dataset.(b) Covariance matrix after feature grouping.}
  \label{fig:feature_grouping_heatmap}
\end{minipage}
\vspace{-8mm}
\end{figure}

\stitle{Feature grouping.}
Naively assuming feature independence ignores the ER feature engineering process
where multiple features may be created for each aligned attribute.
Clearly, each of these features are dependent on each other.
Consider \Cref{fig:feature_grouping_heatmap}(a) that shows the heat map of the correlation matrix of all features for ground-truth matches from the Fodors Zagats dataset (the heat map for unmatches is similar). We can clearly see that some blocks have higher values while other values are closer to 0.
Not surprisingly, each block corresponds to the set of features generated by the same attribute.
This suggests a natural modification to naive GMM:
(1) \textit{features generated from the same attribute are dependent}; and
(2) \textit{ features generated by different attributes are independent}.

Conceptually, the covariance matrix $\Sigma_C$ is now a block diagonal matrix where each block corresponds to the covariance matrix of the features obtained from the same attribute, as shown in \Cref{fig:feature_grouping_heatmap}(b),
where $C \in \{M, U\}$ and $\Sigma_{C}^g$ is the covariance matrix for features of the $g$-th attribute. This results in a dramatically reduced number of parameters.

\stitle{Correlation sharing.}
The traditional approach for parameter sharing in GMM is to assume that covariance values are the same for both M and U distributions,i.e. $\Sigma_M = \Sigma_U$. 
To empirically verify if this can be used in ER, for every dataset, we calculate the sample covariance matrix for matches $S_M$ and unmatches $S_U$ using the ground-truth, and we calculate the cosine similarity between $S_M$ and $S_U$ by flattening both matrices to vectors. If $\Sigma_M = \Sigma_U$, one would expect cosine($S_M$,$S_U$) is close to 1.
As shown in the first row in \Cref{tab:corr_vs_cov}, cosine($S_M$,$S_U$) deviates a lot from 1, which suggests that $\Sigma_M = \Sigma_U$ is not a valid assumption for ER.

\begin{table}[htb]
\renewcommand{\arraystretch}{0.8}
\begin{tabular}{|l|l|l|l|l|l|l|}
\hline
 & FZ & DA & DS & AB & AG \\ \hline
cosine($S_M$,$S_U$) & 0.76 & 0.69 & 0.74 & 0.92 & 0.73 \\ \hline
cosine($R_M$,$R_U$) & 0.97 & 0.94 & 0.98& 0.99 & 0.99 \\ \hline
\end{tabular}
\caption{Cosine similarity between covariance matrix (first row) and correlation matrix (second row) using ground-truth matches and unmatches after feature grouping on five real-world datasets.}
\label{tab:corr_vs_cov}
\vspace{-10mm}
\end{table}

Instead of naively sharing the entire covariance matrix (like prior approaches),
we decompose the covariance matrix and then share only the part
that is consistent over M and U.
Specifically, we decompose $\Sigma_C$ where $C \in \{M, U\}$ based on the well-known relationship between covariance and Pearson correlation of two random variables $A$ and $B$: $\text{correlation}(A, B) = \frac{\text{covariance}(A, B)}{\sigma_A \sigma_B}$,
where $\sigma_A$ and $\sigma_B$ are the standard deviation of $A$ and $B$, respectively.  Correlation can be seen as a normalized measure of covariance -- while $\text{covariance}(A, B) \in [-\infty,\infty]$, $\text{correlation}(A, B) \in [-1,1]$.
We decompose $\Sigma_C$ ($C \in \{M, U\}$)  as follows:
\vspace{-2mm}
\begin{equation}
    \label{eq:covarianceMatrixDecomposition}
    \Sigma_C = \Lambda_C R_C \Lambda_C
\vspace{-2mm}
\end{equation}
where $\Lambda_C$ is a diagonal matrix with $\Lambda_C[j,j]$ represents the standard deviation of feature $j$ in class $C$, and $R_C$ is the Pearson correlation matrix of pairs of features in class $C$.




Our key observation is that, in the ER context, \textit{the Pearson correlation matrices $R_M$ and $R_U$ should be very similar}.
This is because features in a same group are obtained by applying different similarity functions on the same attribute,
so the Pearson correlation between two features reflects
the correlation of their corresponding similarity functions on the same attribute, regardless of matches or unmatches.

To verify our intuition, we compute the cosine similarity between $R_M$ and $R_U$  using ground-truth matches and unmatches. As shown in the second row in \Cref{tab:corr_vs_cov}, $cosine(R_M,R_U)$ is close to 1 consistently across all datasets. This ER specific property allows to rewrite~\Cref{eq:covarianceMatrixDecomposition} as:
\vspace{-2mm}
\begin{equation}
        \label{eq:erCovarianceMatrixDecompositionSmart}
    \Sigma_M = \Lambda_M R \Lambda_M \text{ and } \Sigma_U = \Lambda_U R \Lambda_U
    \vspace{-2mm}
\end{equation}
where $R$ is the shared correlation matrix. Since $R$ is the same for both classes, we can
estimate it using the entire dataset in a preprocessing step and completely eliminate it from the set of parameters $\Theta$.

\stitle{Parameters in the ZeroER generative model.} After feature grouping and correlation sharing, the generative model for ZeroER is parameterized by $\Theta = \{\pi_M, \Lambda_M, \Lambda_U, \mu_M, \mu_U\}$, where $\Lambda_M$ (resp. $\Lambda_U$) is a diagonal matrix and $\Lambda_M[j,j]$ (resp. $\Lambda_U[j,j]$) is the standard deviation of the $j$th feature of class M (resp. class U). Therefore, we have a total number of $4d+1$ parameters, a significant reduction from $d^2 + 3d + 1$ parameters of the naive GMM.

\subsection{Feature Regularization}
\label{sec:feature_regularization}
Another unique challenge of ER is that many features can have extremely small variance and can dominate the computation, effectively rendering other features to have little impact on the predictions. In the extreme case, as shown in Example~\ref{ex:overfitting}, a feature with 0 variance in any component $C \in \{M,U\}$ will make $\det(\Sigma_C) = 0$ (determinant of the covariance matrix), and that feature will solely determine the prediction, known as the singularity problem~\cite{bishop2006pattern}.
\stitle{Existing Approach: Uniform Regularization.}
The traditional approach for addressing the  singularity problem is to add a small constant $\kappa$ to the diagonal entries of $\Sigma_C$, i.e., the variance of every feature. This is known formally as the uniform/Tikhonov regularization~\cite{Duchi_2008, Honorio2013Jul}, in which each feature is regularized uniformly.
This approach is widely used including in the popular sklearn package~\cite{gmm_sklearn}.
While this uniform regularization avoids the singularity problem, i.e., having one feature dictating the predictions, it still does not address the overfitting problem, i.e. some features have much more influence than others. In addition, a constant $\kappa$ that works for one feature might be unsuitable for another.

\begin{figure}[t!]
  \centering
  \centerline{\epsfig{figure=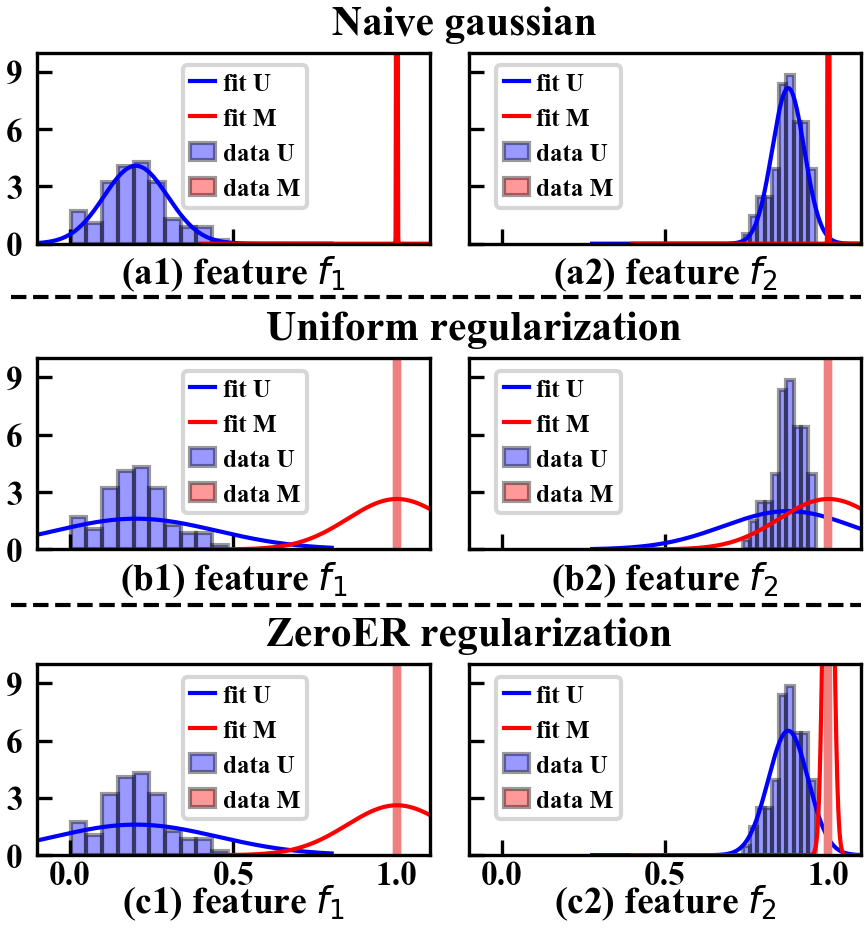, height=8 cm}}
  \vspace{-4mm}
  \caption{Two features $f_1$ and $f_2$ suffer from overfitting problem shown in (a1) and (a2). An uniform regularization works for $f_1$ (b1), but does not work for $f_2$ (b2). The adaptive regularization by \system works for both (c1 and c2).
  }
  \label{fig:singularity}
\vspace{-6mm}
\end{figure}

\vspace{-2mm}
\begin{example}
\label{ex:need_diff_reg}
Consider two features $f_1$ in \Cref{fig:singularity}(a1) and $f_2$ in \Cref{fig:singularity}(a2) that suffer from the overfitting problem. A  $\kappa$ is chosen to appropriately regularize $f_1$ is shown in \Cref{fig:singularity}(b1). The two distributions are now well separated, do not have extremely small variances and $f_1$ could be used for making predictions.
However, applying the same $\kappa$ to $f_2$ results in  \Cref{fig:singularity}(b2), which is clearly an inferior fit. There is too much overlap between the M-distribution and the U-distribution after regularization. This undermines the model's ability to distinguish matches and unmatches.

\vspace{-4mm}
\end{example}

\stitle{Our Solution: Adaptive Regularization.}
Before giving our solution, we first examine how uniform regularization works in principle. Regularization can be seen as adding an ``punishing" term to the objective function. Formally, the uniform regularization modifies the expected data likelihood objective function in the M-Step (c.f. \Cref{eq:m_step}) as:
\begin{equation}
\label{eq:complete_logdatalikelihood_reg}
\mathbb{E}'\{L(\Theta) \} = \mathbb{E}\{L(\Theta) \} - 1/2\times\text{tr}(K (N_M\Sigma_M^{-1}+N_U\Sigma_U^{-1}))
\end{equation}{}
where $K = \kappa I$ is a diagonal matrix acting as regularization parameter; $\text{tr}()$ denotes the trace operation on a matrix -- the sum of all diagonal elements of the matrix; $N_M$ and $N_U$ ensure the regularization term is ``normalized'' to the amount of data $N$. Without it, the first term will dominate when $N$ is large.
Intuitively, this regularization term punishes the variances of M and U being small.

To maximize \Cref{eq:complete_logdatalikelihood_reg}, the optimal values of $\mu_C$ and $\pi_C$ are still given by \Cref{eq:optimal_sigma_mu} but the optimal $\Sigma_C$ is now:
\vspace{-2mm}
\begin{equation}
\label{eq:reg_solution}
    \Sigma_C = S_C + K \text{ where } C \in \{M,U\}
    \vspace{-2mm}
\end{equation}

In other words, designing a regularization strategy is essentially about designing the diagonal regularization matrix $K$.
In uniform regularization, $K$ is chosen to be $K = \kappa I$, so \Cref{eq:reg_solution} adds a number $\kappa$ on every diagonal entry of the covariance matrix. As shown in Example~\ref{ex:need_diff_reg}, using a too large $\kappa$ for feature $f_2$ causes much overlap of the M and U distributions, rendering $f_2$ ineffective in making predictions.

A good regularization needs to meet two goals: (1) the relative influences of dominating features are decreased, so that all features can have a chance to contribute to making predictions; and (2) the more influential features before regularization should still be more influential  after regularization, so as not to decrease the model's predictive capabilities. 
Intuitively, the influence of a feature can be measured by the overlap between its M-Distribution and U-Distribution --- the smaller the overlap, the more influential a feature is. 
We propose to design a diagonal regularization matrix $K = diag(\kappa_1, \kappa_2, \ldots, \kappa_d)$ such that all features have the same amount of \textit{overlap increase}, which would achieve the above two goals. 
To see this, consider a hypothetical scenario where features $f_3$ and $f_4$ have overlaps of $0.001$ and $0.3$, respectively, before regularization, so $f_3$ is dominating $f_4$ in making predictions. After we add the same overlap, say 0.2, $f_3$ and $f_4$ now have overlaps of $0.201$ and $0.5$, respectively.  Clearly, $f_3$ is still more influential than $f_4$, but the relative influences of $f_3$ is decreased significantly (e.g., $\frac{0.3}{0.001} \gg \frac{0.5}{0.201}$).


Specifically, we use Bhattacharyya coefficient (BC) to measure the overlap, as it is a widely accepted measure of the overlap of two distribution and offers a tight upper-bound to Bayes error~\cite{bhattacharyya}, the error rate of a optimal classifier on examples generated by the two distributions.
The Bhattacharyya coefficient of the M and U distribution on the $j$th feature dimension is as follows~\cite{bay_coeff}: 
\vspace{-1mm}
\begin{equation}
\begin{small}
\begin{aligned}
BC_j = \text{exp}\bigg[&-\frac{1}{4}\text{ln}\big(\frac{1}{4}(\frac{\Sigma_M[j,j]}{\Sigma_U[j,j]}+\frac{\Sigma_U[j,j]}{\Sigma_M[j,j]}+2) \big)\\
&-\frac{1}{4}\frac{(\mu_M[j]-\mu_U[j])^2}{\Sigma_M[j,j]+\Sigma_U[j,j]}\bigg]
\end{aligned}
\end{small}
\vspace{-1mm}
\end{equation}


Let $BC_j$ denote the BC coefficient of feature $j$ before regularization, and $BC_j'$ denote the BC coefficient of feature $j$ after regularization, i.e. after adding $\kappa_j$ to the variance of feature $j$ ($\Sigma_C[j,j], C \in \{M, U\}$). For every feature $j$, we would like to design $\kappa_j$ such that $BC_j' - BC_j$ is the same for all features. Let $k' \in [0,1]$ be the regularization parameter that denotes the constant amount of increase of BC in all features. Then we can compute $k_j$ by solving the following equation:
\vspace{-2mm}
\begin{equation}
\label{eq:solve_kappa}
BC_j' - BC_j = \kappa', \forall j \in [1,d]
\vspace{-2mm}
\end{equation}
\Cref{eq:solve_kappa} has only one unknown variable $k_j$ and can be solved efficiently with the newton raphson method~\cite{Weisstein2019Oct}.



\begin{example}
\vspace{-1mm}
Continuing with the previous example on regularizing $f_1$ and $f_2$. Using our adaptive regularization strategy, $f_1$ is fitted in \Cref{fig:singularity}(c1) and $f_2$ is fitted in \Cref{fig:singularity}(c2). As we can see, both features are now well separated and well spread out to avoid the overfitting problem.
\vspace{-1mm}
\end{example}



\vspace{-2mm}
\subsection{ZeroER Algorithm So Far}
We now present the algorithm of \system without transitivity in Algorithm~\ref{alg:zeroer_no_transitivity}. As we can see, the updates to the parameters $\Theta$ happen in the M-Step (Lines~\ref{algo_line:M_start} to~\ref{algo_line:M_end}), and the E-Step re-estimates the posterior probability of every tuple pair (Line~\ref{algo_line:e_step}).
%
%
%
EM algorithm requires an initialization of posterior probabilities $\gamma_{i,C}$ (Line~\ref{algo_line:init_posterior}). To do this, we first use a min-max scaler to normalize every feature into $[0,1]$. We set $\gamma_{i,M}=1$ (hence $\gamma_{i,U}=0$) if $||\textbf{x}_i|| > \epsilon$; otherwise, $\gamma_{i,M}=0$. We choose 0.5 as the default value for $\epsilon$, and show experimentally that \system is robust to the choice of $\epsilon$.


The EM algorithm also requires a termination condition (Line~\ref{algo_line:termination}).
The EM iteration is terminated when the difference of \Cref{eq:m_step} between two consecutive iterations is less than a threshold ($10^{-5}$). We also set a limit the number of EM iterations to 200, a common practice also used in popular ML packages such as sklearn~\cite{gmm_sklearn}. When EM is terminated due to the limit of the maximum number of iterations (instead of likelihood convergence), we average the likelihood results from the latest 20 iterations.



\begin{algorithm}
\caption{ZeroER without transitivity}
\label{alg:zeroer_no_transitivity}
\SetAlgoLined
\KwIn{two tables $T$ and $T'$, a set of similarity functions ${\mathcal F}$, an optional blocking function ${\mathcal B}$}
\KwOut{matching pairs in $T \times T'$}
\label{algo_line:init_start}
$C_s \gets $ all pairs in $T\times T'$ after applying ${\mathcal B}$\\
similarity vector $\textbf{x} \gets$ apply ${\mathcal F}$ on every pair in $C_s$ \\
$R \gets$ the correlation matrix for  feature pairs using all similarity vectors $\textbf{x}'s$. \\
Initialize $\gamma_{i,C}$ ($C \in \{M, U\}$) for all tuple pairs. \label{algo_line:init_posterior}\\
Create variables of \system's generative model $\Theta = \{\pi_M, \Lambda_M, \Lambda_U, \mu_M, \mu_U\}$ that will be updated in the EM loops. \\\label{algo_line:init_end}
\While{Not Converged}{ \label{algo_line:termination}
\textbf{M Step}\\  \label{algo_line:m_step_start}
\Indp  Update $\mu_M$,$\mu_U$,$\pi_M$ by \Cref{eq:optimal_sigma_mu} \label{algo_line:M_start}\\
$\Lambda_C[j,j] = \sqrt{\frac{1}{N_C}\sum_{i=1}^N\gamma_{i,C} (\textbf{x}_i[j]-\Bar{\textbf{x}}_C[j])^2}$ \\
$\Sigma_C \gets \Lambda_C R \Lambda_C$ \\
$K \gets \text{diag}\{\kappa_1,\dots, \kappa_d\}$, where $\kappa_j$ is obtained by solving \Cref{eq:solve_kappa}. \\
$\Sigma_C \gets \Sigma_C + K$ \\ \label{algo_line:M_end}
\Indm
\textbf{E Step}\\
\Indp Update  $\gamma_{i,M}$ by \Cref{eq:e_step}.\\ \label{algo_line:e_step}
\Indm
}
\Return all pairs such that $\gamma_{i,M} > 0.5$
\end{algorithm}
\section{Incorporating Transitivity}
\label{sec:transitivity}



Transitivity constraint stipulates that if
tuple pairs $(t_1, t_2)$ and $(t_1, t_3)$ are matches, then $(t_2, t_3)$
\emph{must} be a match~\cite{DBLP:journals/pvldb/GetoorM12}. While one can enforce transitivity as a postprocessing step, incorporating transitivity into the ER algorithm achieves better performance, as we will verify empirically.

\vspace{-3mm}
\subsection{Transitivity as Posterior Constraint}
\label{sec:transitivity-constraint}
Let $\gamma_{i,j,M}$ denote the posterior probability of a tuple pair $(t_i,t_j)$ being a match, namely $\gamma_{i,j,M} = P(y_{i,j}=M|\textbf{x}_{i,j})$. Note that when we write $\gamma_{i,j,M}$, the two subscripts $i,j$ are used to denote a tuple pair $(t_i,t_j)$; and when we write $\gamma_{i,M}$ (as used in previous sections), the one subscript $i$ is used to denote the $i^{th}$ tuple pair.


Consider three tuple pairs $(t_1, t_2)$, $(t_1, t_3)$, and $(t_2, t_3)$.
Transitivity property~\cite{DBLP:journals/pvldb/GetoorM12} states that if both $(t_1, t_2)$ and $(t_1, t_3)$ are matches, then $(t_2, t_3)$ must be match.
However, it is possible that $(t_2, t_3)$ is a match, but not both of $(t_1, t_2)$ and $(t_1, t_3)$.
We capture this observation as a probabilistic inequality that needs to be satisfied:
\vspace{-2mm}
\begin{equation}
\label{eq:transitivity}
    \gamma_{1,2,M} \times \gamma_{1,3,M} \leq \gamma_{2,3,M}
\vspace{-2mm}
\end{equation}
\Cref{eq:transitivity} secures the transitivity property.
Consider a deterministic scenario that $(t_1, t_2)$ and $(t_1, t_3)$ are known to be matches so that $\gamma_{1,2,M}=\gamma_{1,3,M}=1$. By \Cref{eq:transitivity} we have $\gamma_{2,3,M}\geq 1$, so $(t_2, t_3)$ must also be a match.
Now consider a probabilistic scenario that $\gamma_{1,2,M}=0.7$ and $\gamma_{1,3,M}=0.6$, which means in $70\%$ of the cases $(t_1, t_2)$ is a match and in $60\%$ of the cases $(t_1, t_3)$ is a match. Transitivity occurs when $(t_1,t_2)$ and $(t_1,t_3)$ are both matches which is
$70\% \times 60\% = 42\%$ of the cases. Therefore, $(t_2,t_3)$ has at least $42\%$ chances of being a match due to transitivity, as captured by \Cref{eq:transitivity}.

Let $\boldsymbol{\gamma}$ denote the vector of posterior probabilities of being a match for all tuple pairs in $T \times T'$. For each trio of tuples $t_i, t_j, t_k$, we can define the probabilistic constraint
from \Cref{eq:transitivity}.
The set of all such constraints define a feasibility set for $\boldsymbol{\gamma}$:
\vspace{-2mm}
\begin{equation}
\label{eq:feasibility_set_Q}
Q = \{\boldsymbol{\gamma}|\gamma_{i,j,M}\gamma_{i,k,M} - \gamma_{j,k,M} \leq 0\  \forall i,j,k \}
\vspace{-1mm}
\end{equation}

Recall from \Cref{sec:generative_model}, the objective function of the generative model is data likelihood function $L(\Theta)$ parameterized by $\Theta$, which is maximized via an EM algorithm.
The posterior probabilities $\boldsymbol{\gamma}$ are computed based on $\Theta$ via Bayes rule (c.f. \Cref{eq:bayesRulePosteriorGeneral}).
On the surface, it seems very difficult to incorporate $Q$ in the learning process, namely, finding the best $\Theta$ that maximizes $L(\Theta)$, while ensuring the posterior probabilities $\boldsymbol{\gamma}$ computed based on $\Theta$ satisfies $Q$.

\stitle{Incorporating Posterior Constraints in EM.}
Fortunately, this conundrum could be solved by using the \emph{free energy} view of the EM algorithm.
In this view, the EM algorithm is seen to optimize a different objective function, the free energy function, that treats both $\Theta$ and $\boldsymbol{\gamma}$ as its parameters.
Previously, free energy based approach was proposed to develop the incremental and sparse variants of the EM algorithm~\cite{neal1998view}. We adopt this approach as it allows us incorporate the transitivity constraints directly into the optimization process.  The free energy function $F(\Theta, \boldsymbol{\gamma)}$~\cite{neal1998view,jordan2004introduction} is defined as:
\begin{small}
\begin{equation}
\label{eq:EM_obj_maxE}
\begin{aligned}
F(\Theta, \boldsymbol{\gamma})=&\Sigma_{i=1}^n\Sigma_{C\in\{M,U\}}\big[ \gamma_{i,C}\text{log } \big(\pi_C p(\textbf{x}_i|\theta_C)/\gamma_{i,C}\big)\big]
\end{aligned}
\end{equation}
\end{small}
It turns out $F(\Theta, \boldsymbol{\gamma})$ can be optimized the exact same way as $L(\Theta)$ using the same EM algorithm, with a different view of the E-Step and the M-Step~\cite{neal1998view}. Specifically, the E-Step is now seen as maximizing $F(\Theta, \boldsymbol{\gamma})$ with respect to $\boldsymbol{\gamma}$ given  $\Theta^{t-1}$ estimated from the previous iteration:
\vspace{-1mm}
  \begin{equation}
  \label{eq:e_step_maxE}
 \boldsymbol{\gamma}^t = \text{arg}\max_{\boldsymbol{\gamma}} F(\Theta^{t-1}, \boldsymbol{\gamma})
 \vspace{-1mm}
  \end{equation}
and the M-Step is now seen as maximizing $F(\Theta, \boldsymbol{\gamma})$ with respect to $\Theta$ given  $\boldsymbol{\gamma}^{t-1}$ estimated from the previous iteration:
\vspace{-2mm}
  \begin{align}
  \label{eq:m_step_maxE}
  \Theta^t = \text{arg}\max_{\Theta} F(\Theta, \boldsymbol{\gamma}^{t})
  \vspace{-1mm}
  \end{align}
Once we adopt this free energy view of EM, the constraints in \Cref{eq:feasibility_set_Q} can be incorporated by adapting the E-step as a constrained optimization problem, namely:
\vspace{-1mm}
\begin{equation}
\label{eq:e_step_constraint}
\boldsymbol{\gamma}^t = \text{arg}\max_{\boldsymbol{\gamma} \in Q} F(\Theta^{t-1}, \boldsymbol{\gamma})
\vspace{-1mm}
\end{equation}
\begin{lemma}
\label{lemma:constrained_max_non_convex}
The constrained maximization problem in \Cref{eq:e_step_constraint} is a non-convex optimization problem.
\end{lemma}
%
We leave the full proof of Lemma~\ref{lemma:constrained_max_non_convex} to the full technical report~\cite{tech_report}.
Typically, heuristic methods such as projected gradient descent and alternating minimization
are used to obtain an approximate solution for the non-convex optimization~\cite{jain2017non, tuy1988global}.
Unfortunately, these methods involve an additional expensive iterative process at every E-step drastically increasing the overall running time.
We introduce a much more efficient solution next.

\subsection{Solving the Constrained Problem}
\label{sec:transitivity-solution}

\begin{figure}[htb]
\vspace{-2mm}
\begin{minipage}[b]{1.0\linewidth}
  \centering
  \centerline{\epsfig{figure=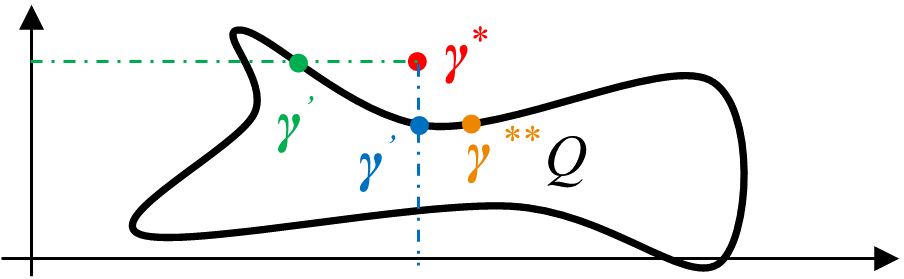,width=5cm}}
  \vspace{-3mm}
  \caption{Illustrations of different solutions to the constrained maximization problem in 2-D space.}
  \label{fig:projection_inuition}
\end{minipage}
\vspace{-7mm}
\end{figure}

Let $\boldsymbol{\gamma}^*$ be the optimal solution for the unconstrained optimization problem in  \Cref{eq:e_step_maxE}, which is given by \Cref{eq:e_step}.
Let $\boldsymbol{\gamma}^{**}$ be the unknown optimal solution for the constrained optimization problem in \Cref{eq:e_step_constraint}.
If $\boldsymbol{\gamma}^*$ already satisfies $Q$, then obviously $\boldsymbol{\gamma}^{**} = \boldsymbol{\gamma}^*$; otherwise, we need an efficient way of computing $\boldsymbol{\gamma}^{**}$.
Though the optimization problem in \Cref{eq:e_step_constraint} is non-convex (due to non-convex constraints $Q$), the objective function $F(\Theta,\boldsymbol{\gamma})$ is convex with respect to $\boldsymbol{\gamma}$.
Therefore, we know that the solution $\boldsymbol{\gamma}^{**}$ has to be on the boundary of $Q$~\cite{boyd2004convex}. 
\Cref{fig:projection_inuition} shows different solutions in 2D, where $\boldsymbol{\gamma}^*$ is outside of $Q$ and $\boldsymbol{\gamma}^{**}$ is on $Q$.


As discussed, finding the exact $\boldsymbol{\gamma}^{**}$ is expensive.
However, it is possible to \emph{approximate} $\boldsymbol{\gamma}^{**}$ by $\boldsymbol{\gamma}^{\prime}$ on the boundary of $Q$ that can be found much more efficiently. Our intuition is that the closer the approximate solution $\boldsymbol{\gamma}^{\prime}$ is to the known optimal solution $\boldsymbol{\gamma}^{*}$ of the unconstrained problem, the better the approximation is.

The operation of finding $\boldsymbol{\gamma}^{\prime}$ based on its distance with $\boldsymbol{\gamma}^{*}$ is known as projection~\cite{jain2017non}. Since there are many different ways to measure the "closeness" between $\boldsymbol{\gamma}^{\prime}$ and $\boldsymbol{\gamma}^{*}$, there are many different ways of doing the projection. We propose an efficient way of projection that requires minimum changes on $\boldsymbol{\gamma}^{*}$, and at the same time, tries to maximize $F(\Theta,\boldsymbol{\gamma})$.

\stitle{Handling one constraint.}
Consider a single constraint in $Q$: $\gamma_{i,j,M}\times \gamma_{i,k,M} \leq \gamma_{j,k,M}$.
When the inequality is violated, one can adjust one of the three probabilities to make it hold: decrease $\gamma_{j,k,M}$, increase $\gamma_{i,j,M}$ or increase $\gamma_{i,k,M}$ so that $\gamma_{i,j,M}\times \gamma_{i,k,M} = \gamma_{j,k,M}$.
In \Cref{fig:projection_inuition}, these adjustments correspond to projecting $\boldsymbol{\gamma}^*$ on the boundary of $Q$ along the axises to be the green or blue points.
Formally, the three ways of projection along the three axis are:
\begin{small}
\begin{equation}
\label{eq:projection_along_axis}
\gamma_{i,j,M}' = \frac{\gamma_{j,k,M}}{\gamma_{i,k,M}},\
\gamma_{i,k,M}' = \frac{\gamma_{j,k,M}}{\gamma_{i,j,M}},\  \gamma_{j,k,M}' = \gamma_{i,j,M}\gamma_{i,k,M}
\end{equation}
\end{small}
Therefore, given the optimal solution $\boldsymbol{\gamma}^*$ to \Cref{eq:e_step_maxE} that does not satisfy the one constraint in $Q$, we obtain three possible approximate solutions by updating each one of the three posterior probabilities as in \Cref{eq:projection_along_axis}, and we pick the one
that has the maximal $F(\Theta,\boldsymbol{\gamma})$.




\stitle{Handling multiple constraints.}
Now we consider multiple constraints $Q=\{g_1, g_2,...\}$. 
A natural approach is to do the projection in \Cref{eq:projection_along_axis} for each of the constraints. However, a probability $\gamma_{i,j,M}$ might get involved in several constraints. For example, $\gamma_{i,j,M}$ is involved in two constraints $\{g_1, g_2\}$, specifically the following two cases can happen:
1) The projection by \Cref{eq:projection_along_axis} in constraint $g_1$ has increased/decreased $\gamma_{i,j,M}$ to $\gamma_{i,j,M}'$, and the projection in the current constraint $g_2$ requires $\gamma_{i,j,M}'$ to be decreased/increased, which would make $g_1$ unsatisfied. In this case, we adopt a greedy approach: When the best projection axis in \Cref{eq:projection_along_axis} causes conflicts with a previous constraint, we project along the second best axis. If it causes conflicts again we project along the third axis and if conflicts happen again we perform no projection on this constraint.
2) Constraint $g_1$ increased/decreased $\gamma_{i,j,M}$ to $\gamma_{i,j,M}'$. After adjusting $\gamma_{i,j,M}$ to be $\gamma_{i,j,M}'$, constraint $g_2$ also becomes satisfied. Or, constraint $g_2$ requires $\gamma_{i,j,M}'$ to be increased/decreased further. In this case, we just increase/decrease $\gamma_{i,j,M}'$ further.

Furthermore, there are two cases when a probability $\gamma_{i,j,M}$ involved in the constraints doesn't exist in the posterior vector $\boldsymbol{\gamma}^*$. First, the corresponding tuple pair $(t_i, t_j)$ was excluded by blocking; in this case, we assume $\gamma_{i,j,M} = 0$. Second, $\gamma_{i,j,M}$ is not in $\boldsymbol{\gamma}^*$ when $t_i$ and $t_j$ are from the same table, as $C_s$ only contains cross-table tuples pairs. This is not a problem when $T = T'$ because $(t_i, t_j)$ is a cross-table tuple pair at the same time. We handle  the case when $T\neq T'$ in \Cref{sec:transitivity_2_tables}.

\stitle{Reducing the number of constraints.}
The efficiency of our approach is reliant on the number of constraints that are checked in each iteration, which is cubic w.r.t. the number of tuples.
However, it is possible to prune most of these constraints.
Specifically, transitivity property is only useful for tuple pairs that are predicted to be matches.
Since the number of matches are orders of magnitude smaller than unmatches, the number of constraints also reduce significantly. Specifically, we only need to consider the following reduced set of constraints:
\vspace{-3mm}
\begin{small}
\begin{equation}
\label{eq:relaxed_Q}
Q' =  \{\boldsymbol{\gamma}|\gamma_{i,j,M}\gamma_{i,k,M} \leq \gamma_{j,k,M},  \text{where }\gamma_{i,j,M} \geq 0.5, \gamma_{i,k,M} \geq 0.5  \}
\end{equation}
\end{small}
\vspace{-8mm}
\subsection{Transitivity Across Two Tables}
\label{sec:transitivity_2_tables}
Incorporating transitivity in the general scenario when $T \neq T'$ is more involved.
In \Cref{eq:transitivity}, the three involved tuple pairs are $(t_1, t_2)$, $(t_1, t_3)$, and $(t_2, t_3)$; two of the three tuples $t_1, t_2, t_3$ must come from the same table, and the third tuple must come from the other table. Without loss of generality, let us assume $t_2, t_3$ are from the same table. Using one generative model for $T \times T'$ discussed before only gives the values of $\gamma_{1,2,M}$ and $\gamma_{1,3,M}$ but not $\gamma_{2,3,M}$.
If we know both $T$ and $T'$ are duplicate-free, we can set $\gamma_{2,3,M} = 0$; however, we make no such assumption in this work.

One simple way to be able to compute the posterior probabilities of tuple pairs in the same table is to use the same generative model parameterized by $\Theta$, and include in the data likelihood function $L(\Theta)$ (\Cref{eq:complete_logdatalikelihood}) not only the $ N$ tuple pairs in $T \times T'$, but also the $N_l$ tuple pairs in $T \times T$ and the $N_r$ tuple pairs in $T' \times T'$. However, this approach produces inferior results primarily because the distribution of similarities vectors for tuple pairs in $T \times T'$ is quite different from those in $T \times T$ or in $T' \times T'$. For example, as shown in \Cref{fig:ER_illu}, the phone area code is separated by ``/" in the fodors table while by ``-" in the zagats table, and city name ``los angeles" is written with full name in fodors while in zagats it is abbreviated as ``la".

Given the above observation, we propose to use a generative model with three sets of parameters: the original set $\Theta$ for capturing similarity vectors of tuple pairs in $T \times T'$, and two new sets $\Theta_l$ and $\Theta_r$ for capturing similarity vectors of tuple pairs in $T \times T$ and $T' \times T'$, respectively. The new data likelihood function becomes:
\vspace{-1mm}
\begin{small}
\begin{align}
\label{eq:complete_logdatalikelihood_3_paras}
    L(\Theta,\Theta_l, \Theta_r) =& \Sigma_{i=1}^{N} \Sigma_{C \in \{M,U\}} \mathbbm{1}_{y_i = C}\log \big( \pi_{C} p(\textbf{x}_i|\theta_{C}) \big) \nonumber \\
    +\Sigma_{i=N+1}^{N+N_l}& \Sigma_{C \in \{M,U\}} \mathbbm{1}_{y_i = C}\log \big( \pi_{l,C} p(\textbf{x}_i|\theta_{l,C}) \big) \nonumber \\
    +\Sigma_{i=N+N_l+1}^{N+N_l+N_r}& \Sigma_{C \in \{M,U\}} \mathbbm{1}_{y_i = C}\log \big( \pi_{r,C} p(\textbf{x}_i|\theta_{r,C}) \big)
\end{align}
\vspace{-1mm}
\end{small}
\Cref{eq:complete_logdatalikelihood_3_paras} is essentially the sum of \Cref{eq:complete_logdatalikelihood} with three different set of parameters:
\vspace{-2mm}
\begin{equation}
L(\Theta_l,\Theta_r, \Theta) = L(\Theta_l) + L(\Theta_r) + L(\Theta)
\vspace{-2mm}
\end{equation}
Similarly, we can view the data likelihood function as the new free energy function that treats $\boldsymbol{\gamma}_l$, $\boldsymbol{\gamma}_r$, and $\boldsymbol{\gamma}$ as parameters:
\begin{equation}
\label{eq:big_free_energy}
F(\Theta_l, \boldsymbol{\gamma}_l,\Theta_r,\boldsymbol{\gamma}_r,\Theta, \boldsymbol{\gamma}) = F(\Theta_l, \boldsymbol{\gamma}_l)+F(\Theta_r, \boldsymbol{\gamma}_r)+F(\Theta, \boldsymbol{\gamma})
\end{equation}
where $F(\Theta_l, \boldsymbol{\gamma}_l)$, $F(\Theta_r, \boldsymbol{\gamma}_r)$ and $F(\Theta, \boldsymbol{\gamma})$ are the three free energy functions for the three types of similarity vectors.

Note that we still have one generative model with three sets of parameters, instead of three independent generative models, as the three sets of parameters are linked by transitivity constraints. Algorithm~\ref{alg:zeroer} describes the final \system procedure incorporating transitivity. As we can see, the majority of the procedure is about running the corresponding part in Algorithm~\ref{alg:zeroer_no_transitivity} three times for $T \times T'$, $T \times T$, and $T' \times T'$, including the initialization, the M-Step, and the updating posterior probabilities part of the E-Step. The only major change is in Line~\ref{algo_line:transitivity} that resolves transitivity violations once we have updated posterior probabilities using the method described in Section~\ref{sec:transitivity-solution}.



\begin{algorithm}
\setstretch{0.9}
\caption{ZeroER}
\label{alg:zeroer}
\SetAlgoLined
\KwIn{two tables $T$ and $T'$, a set of similarity functions ${\mathcal F}$, an optional blocking method ${\mathcal B}$}
\KwOut{matching pairs in $T \times T'$}
For each of $(T,T')$, $(T,T)$ and $(T',T')$,
execute Line~\ref{algo_line:init_start} to Line~\ref{algo_line:init_end} in Algorithm~\ref{alg:zeroer_no_transitivity} \\
\While{Not Converged}{
\textbf{M Step}\\
\Indp Run Line~\ref{algo_line:M_start} to Line~\ref{algo_line:M_end} in Algorithm~\ref{alg:zeroer_no_transitivity} for left,right and cross table to update the three set of parameters $\Theta_l$, $\Theta_r$ and $\Theta$.\\
\Indm
\textbf{E Step}\\
\Indp Obtain $\boldsymbol{\gamma}_l^*$,  $\boldsymbol{\gamma}_r^*$ and  $\boldsymbol{\gamma}^*$ using \Cref{eq:e_step} with $\Theta_l$, $\Theta_r$ and $\Theta$, respectively.\\ 
Resolve transitivity constraints, namely, update $\boldsymbol{\gamma}_l$,  $\boldsymbol{\gamma}_r$ and  $\boldsymbol{\gamma}$, based on $\boldsymbol{\gamma}_l^*$,  $\boldsymbol{\gamma}_r^*$ and  $\boldsymbol{\gamma}^*$, using the method in \Cref{sec:transitivity-solution}\\ \label{algo_line:transitivity}
\Indm
}
\Return all pairs corresponding to entries in $\boldsymbol{\gamma}^*$ that are greater than 0.5

\end{algorithm}

\section{Experiments}
\label{sec:exp}


We conduct an extensive set of experiments to evaluate the efficacy of ZeroER along three dimensions:
\begin{itemize}[wide,labelindent=0pt]
    \itemsep0em
    \item \emph{Feasibility and Performance of ZeroER (Section~\ref{subsec:autoer_performance})}
        Is it possible for an unsupervised method to
        achieve performance comparable to supervised ML algorithms? How does it compare with existing unsupervised methods?
    \item \emph{Ablation Analysis (Section~\ref{ssec:ablation}).}
        How does our innovations (generative modelling, feature regularization, and incorporating transitivity) in \system contribute to its accuracy?
    \item \emph{Sensitivity Analysis (Section~\ref{ssec:sensitivity_analysis}).}
        Is ZeroER sensitive to the size of the dataset, the regularization hyperparameter,  its initialization, the set of features used, and the strength of the blocking function used?
\end{itemize}

\subsection{Experimental Setup}
\label{subsec:expSetup}
\textbf{Hardware and Platform.}
All our experiments were performed on a machine with a 2.20GHz Intel Xeon(R) Gold 5120 CPU, \revisenew{a Tesla K80 GPU} and with 96GB 2666MHz RAM. 


\stitle{Datasets.}
\revisenew{We focus on the ER task of structured datasets.} We conducted extensive experiments on five \revisenew{structured} datasets from three diverse domains including publication, e-commerce and restaurants.
\Cref{tbl:datasetCharacteristics} provides statistics of these datasets.
All are popular benchmark datasets  extensively used in prior ER work.

\begin{table}[h!]
\vspace{-4mm}
\begin{center}
\setlength{\tabcolsep}{0.1em}
\renewcommand{\arraystretch}{0.8}
\begin{tabular}{|c|c|c|c|}
    \hline
    {\bf Dataset} & {\bf \#Tuples} & {\bf \#Ms} & {\bf \#As}\\\hline
    FZ, Fodors-Zagat~\cite{uci}& 533 - 331 & 112 & 7 \\ \hline
    DA, DBLP-ACM~\cite{erhardws} & 2,616 - 2,294 & 2,224 & 4 \\ \hline
    DS, DBLP-Scholar~\cite{erhardws} & 2,616 - 64,263 & 5,347 & 4 \\ \hline
    AB, Abt-Buy~\cite{erhardws} &1,082 - 1,093 &1,098 &3  \\ \hline
    AG, Amazon-Google products~\cite{erhardws} &1,363 - 3,226 &1,300 &4  \\ \hline
\end{tabular}
\caption{Datasets characteristics, Ms and As denote matches and attributes}
\label{tbl:datasetCharacteristics}
\end{center}
\vspace{-12mm}
\end{table}


\stitle{Algorithms Evaluated.}
We compare ZeroER against 4 supervised methods (1-4) and 5 unsupervised methods (5-9), and 1 active learning method (10).
The ten algorithms are:
\begin{enumerate}[wide,labelindent=0pt]
    \itemsep0em
    \item \emph{Logistic Regression (LR):} This is a typical linear classifier. We use 5-fold cross validation to tune the $\ell_2$ regularization parameter.
    \item \emph{Random Forest (RF):} This is a typical tree-based classifier. The number of trees is set as 100 and the minimum number of samples required to be at a leaf node is tuned by a 5-fold cross validation to avoid overfitting.
    \item \emph{Multi-layer Perceptron (MLP):} This is a typical deep learning classifier. We use two hidden layers with 50 and 10 hidden units and tuned $\ell_2$ regularization parameter by 5-fold cross validation.

	\item \revisenew{\emph{DeepMatcher (DM):} DeepMatcher~\cite{anhaisigmod2018} with hybrid attribute summarization is the state-of-the-art deep learning method for entity resolution. We use the open-source implementation~\cite{anhaidgroup2019Dec} from the authors. Hyperparameters are tuned by grid search and a validation set is used to select the best model snapshot among all epochs to prevent overfitting~\cite{anhaisigmod2018}.}
    
    \item \emph{K-Means (KM-SK):}
        This baseline applies the K-Means algorithm from Scikit-Learn with $K=2$.
        If the features vectors for matches and unmatches are very different,
        then this method would provide good results.
    \item \emph{K-Means (KM-RL):}
        This is an improved baseline from~\cite{RecordLinkageGH} that is calibrated for two-cluster ER task.
        Traditional K-Means often fails when the sizes of two clusters are very uneven~\cite{KMeansIssues}
        which is often the case in ER.
        This algorithm tackles the class imbalance through class weighting so that matches get a higher weight than unmatches.
    \item \emph{GMM:}
        This applies the naive Gaussian Mixture Model from scikit-learn with $2$ components.
    \item \emph{ECM:}
        The Fellegi-Sunter (FS) model~\cite{fellegi1969theory} is a seminal approach for probabilistic unsupervised ER that uses binary features and assumes feature independence.
        We used an improved  implementation from~\cite{RecordLinkageGH,de2015probabilistic} that provides better results than the initial FS model. This approach implements a expectation conditional maximization algorithm that relaxes the simplistic feature independence assumption from FS.
     \item \revisenew{\emph{PPjoin:} This is a similarity-join method that employs filtering techniques to improve computational efficiency~\cite{xiao2011efficient}.  Since PPJoin is a single-attribute joining algorithm, we apply PPJoin by concatenating all attributes. 
     PPJoin implemented and evaluated~\cite{xiao2011efficient} optimized version of two similarity functions (Jaccard and Cosine). 
     PPjoin also requires a threshold to be set. For each dataset, we search the threshold from 0 to 1 by a step size of 0.2 and try both similarity function and report the best result (denoted by PP$^*$).}
     
     \item \emph{Active learning based Random forest (AL-RF):} We use the implementation from modAL~\cite{modAL2018} and use the default query strategy (uncertainty sampling).
\end{enumerate}

\vspace{-3mm}
\stitle{Setups for Various Algorithms.} \revisenew{In this work, we treat feature engineering as a black-box}, the default set of features for each dataset is generated by Magellan~\cite{konda2016magellan}, which are used by all methods \revisenew{except DM that uses embedding features and PPjoin that uses its few supported similarity functions.} 
Under the hood, Magellan infers a type for each aligned attribute, and applies a set of pre-defined similarity functions for each type to generate features. All possible attribute types and the similarity functions considered for each attribute type can be found at~\cite{py_em_doc}. \revisenew{We also design a default blocking strategy for each dataset based on its most informative features (e.g., title) using locality sensitive hashing; the same blocking is again employed by all compared methods.}
For all methods that need a seed, the reported results are the average results of ten runs. We will conduct sensitivity analysis using different features and blocking strategies in Section~\ref{ssec:sensitivity_analysis}.

We use the sklearn~\cite{scikit-learn} implementation for \revisenew{LR, RF, and MLP}. We randomly split each dataset to training and test set by 50\%-50\%. Note that using 50\% of the data as training data is very generous as in practice the amount of labeled data available is much smaller. The match entries in the training set are over-sampled as is typically done in training supervised ML methods in the presence of class imbalance. \revisenew{For DM, we hold out $\frac{1}{3}$ of the training set as validation set.}
\revisenew{For all unsupervised methods, we use the whole dataset for fitting without train-test split.}
For active learning, it starts with 10 random examples and queries one example at every learning iteration and is terminated once it has queried 50\% of the matches or 50\% of all examples; We evaluate the performance on the remaining examples.
%
%
%
%
%
By default, we set the feature regularization parameter $\kappa'$ for ZeroER to $1\%$; we set the initialization threshold $\epsilon$ to be $0.5$; 
We evaluate the sensitivity of \system to these parameters in Section~\ref{ssec:sensitivity_analysis}.

\vspace{-1mm}
\stitle{Performance Measures.}
We used F-score as the performance measure as ER is a task with unbalanced labels.


\vspace{-2mm}
\subsection{ZeroER Performance}
\label{subsec:autoer_performance}

\textbf{Overall comparison.} \Cref{tbl:benchmark_all} reports the results.
\begin{itemize}[wide,labelindent=0pt]
    \itemsep0em
    \item \emph{ZeroER vs unsupervised methods:}  As we can see, though \system builds upon GMM, \system  significantly outperforms GMM. ECM is simply not competitive as it binarizes  all features and uses Bernoulli mixture model, which loses information. Out of the two K-Means clustering algorithms (KM-RL and KM-SK), KM-RL is obviously better as it addresses the cluster imbalance problem observed in ER tasks. \revisenew{ZeroER is on average $11\%$ better than PP$^*$(PPjoin with the best configuration obtained using the ground-truth) on every dataset. Note that in practice it is not possible to arrive at the best configuration for PPJoin without sufficient labeled data. We also observed that the performance of PPjoin is very sensitive to the threshold, and its F-score can decrease by as large as $0.67$ with a $0.2$ change in the threshold.} Overall, \system greatly outperforms all five unsupervised methods due to the ER-specific innovations.




    \item \emph{ZeroER vs supervised methods:}
    The performance of ZeroER is comparable to those supervised methods. This is really impressive as they use the same set of features; while  supervised methods have 50\% of all data as training, \system is completely unsupervised. \revisenew{DM shows better results on AB and AG because these two datasets contain large strings. However, as shown in \Cref{tbl:supervised_benchmark}, DM requires more than ten thousand labels to achieve this performance.}
\item \emph{ZeroER vs active learning:} Active learning is able to even outperform supervised learning. This is because we give active learning the same labeling budget as supervised learning (50\% of labeled data), and active learning is able to judiciously select which 50\% data to use. In practice, though, active learning budget tends to be much lower.  \revisenew{ZeroER is much worse than active learning on DS, because ZeroER relies on a hypothesis that may be satisfied to different degrees on different datasets. The hypothesis is that the matches and unmatches can be divided into two clusters under gaussian distributions.}


\end{itemize}


\begin{table}[htb]
\vspace{-5mm}
\small
\begin{center}
\setlength{\tabcolsep}{0.1em}
\scalebox{0.85}{
\renewcommand{\arraystretch}{0.9}
\begin{tabular}{|c|c|ccccc|cccc|c|}
\hline
&\multicolumn{1}{c|}{}&\multicolumn{5}{c|}{\bf{Unsupervised}}&\multicolumn{4}{c|}{\bf{Supervised}}&\multicolumn{1}{c|}{\bf{\makecell{Active\\ learning}}}\\ \hline
              &ZeroER& ECM &KM-RL & KM-SK & GMM&\revisenew{PP$^*$} &RF &LR &MLP&\revisenew{DM} &AL-RF\\ \hline
FZ &\bf{1} & 0.07   & 0.30   & 0.30  &0.30 &\revisenew{0.97} &0.97 &0.98 &0.99&\revisenew{0.93} & \bf{1} \\\hline
DA     &0.96 & 0.09   & 0.95  & 0.27   &0.26&\revisenew{0.87}  &0.98 &0.96 &0.97&\revisenew{0.97}& \bf{0.99} \\\hline
DS  &0.86 & 0.07   & 0.85	&0.43   &0.07 &\revisenew{0.83} &0.93 &0.88&0.92&\revisenew{0.95} & \bf{0.99} \\\hline
AB      &0.52 & 0.01   &0.01    &0.02	&0.02&\revisenew{0.29} &0.46&0.18 &0.32&\revisenew{\bf{0.63}}   & 0.44 \\\hline
AG     &0.48  & 0.01   &0.02    &0.02	&0.02&\revisenew{0.30} &0.51 &0.18&0.35&\revisenew{\bf{0.67}}  & 0.46 \\\hline
average & 0.76 & 0.05&0.43&0.19&0.13&\revisenew{0.65} &0.77&0.64&0.71&\revisenew{\bf{0.83}}&0.78\\\hline
\end{tabular}
}
\caption{F-score for all methods.}
\label{tbl:benchmark_all}
\end{center}
\vspace{-9mm}
\end{table}

\begin{figure}[htb]
 \vspace{-3mm}
  \centering
  \centerline{\epsfig{figure=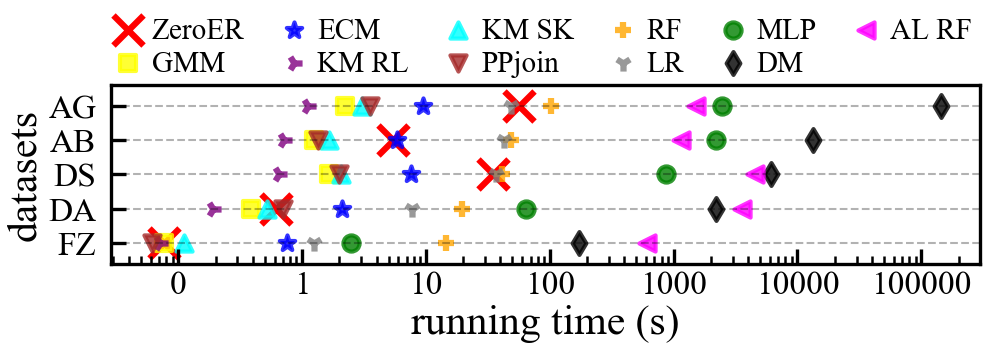, width=8.5cm}}
  \vspace{-4mm}
  \caption{\revisenew{Running time for all algorithms.}}
  \label{fig:running_time}
\vspace{-5mm}
\end{figure}

\noindent \textbf{Overall running time comparison.} \Cref{fig:running_time} shows that the running time of ZeroER is comparable to other methods. Active learning and \revisenew{DM} are typically 10 to 100 times slower than other methods. \revisenew{Active learning will be even slower in practice because we only reported machine time and human time should also be considered in practice.} 
\revisenew{The relative running time of ZeroER varies across datasets. This is because the running time of ZeroER is affected by the number of matches (\Cref{eq:relaxed_Q}) and the time for solving the regularization parameter (\Cref{eq:solve_kappa}) and these two vary across different datasets.} \revisenew{Note the reported time in \Cref{fig:running_time} is for matching only, and does not include the time for blocking and feature generation.}

\begin{table}[th!]
\begin{center}
\renewcommand{\arraystretch}{0.8}
\scalebox{0.85}{
\begin{tabular}{|c|c|c|c|c|c|c|}
    \hline
     & { LR} & { RF} & { MLP}&{ \revisenew{DM}}&{ AL-RF} \\ \hline
    FZ & $2915^*$ & $2915^*$ & $2915^*$ &\revisenew{2332}&1572 \\ \hline
    DA& 418 & 232&417&\revisenew{4647}&26 \\ \hline
    DS& 413 & 227&270&\revisenew{6768}&33  \\ \hline
    AB& $162981^*$  & $162981^*$&$162981^*$& \revisenew{16865} &$162981^*$ \\ \hline
    AG & $358281^*$ & 7589&$358281^*$&\revisenew{17916} &$358281^*$ \\ \hline
\end{tabular}
}
\caption{\small{\# labeled examples needed for supervised and active learning methods to match ZeroER's F-score. Numbers with an asterisk denote all tuple pairs in a dataset since those methods are not able to match ZeroER's performance.}}
\label{tbl:supervised_benchmark}
\end{center}
\vspace{-8mm}
\end{table}

\noindent \textbf{Labeling effort saved.}
We further investigate how much labeled data does the supervised and active learning methods need
to match the performance of ZeroER. Table~\ref{tbl:supervised_benchmark} shows the labeling efforts saved is quite substantial. Active learning in general is able to reduce the labeled examples required compared with supervised learning, though it requires human involvement as the algorithm runs. \revisenew{Though, DM achieves the best average performance, it requires thousands of labels.
On easy datasets (FZ, DA and DS), DM can require up to 30 times more labels than traditional supervised methods.}

\subsection{Ablation Analysis. }
\label{ssec:ablation}

%

We next perform a series of experiments to understand the contributions of
different components of ZeroER and how do they compare with existing solutions to address the same problem. Results are summarized in  \Cref{tab:ablation}. \revisenew{The right-most three columns are the results of replacing each of the three \system innovations (for data deficiency, feature regularization and transitivity) to the naive ones respectively.} 

\begin{table}[h!]
\vspace{-4mm}
\setlength{\tabcolsep}{0.1em}
\small
\begin{center}
\scalebox{0.9}{
\renewcommand{\arraystretch}{0.85}
\begin{tabular}{|=c|+c|+c|+c|+c|}
\hline
&\makecell{\system}&\makecell{group+share corr\\$\to$ diag+share cov } &\makecell{adaptive reg\\$\to$ uniform reg} &\makecell{posterior constraint\\ $\to$ post-processing}\\ \hline
FZ &\textbf{1} &0.97 &0.95&0.99 \\ \hline
DA &0.96 & 0.96  &0.36 & \textbf{0.97}   \\ \hline
DS &\textbf{0.86}& 0.78 &0.59	&0.41   \\ \hline
AB &\textbf{0.52}&0.08 &0.07   &0.45	  \\ \hline
AG &\textbf{0.48} &0.09 &0.04   &0.42	 \\ \hline
average&\textbf{0.76}&0.57&0.38&0.65 \\ \hline
\end{tabular}
}
\caption{\revisenew{Ablation analysis for \system.}}
\label{tab:ablation}
\end{center}
\vspace{-10mm}
\end{table}


\stitle{Handling data deficiency.} Our method of handling data deficiency (feature grouping and sharing correlation matrix) works better than existing commonly used approaches (assuming feature independence and sharing covariance matrix) on \revisenew{four} datasets (FZ, \revisenew{DS,} AB and AG).

\vspace{-2mm}
\stitle{Feature regularization.} We compare our adaptive regularization to uniform regularization. Since there is no magic constant that works for all features in uniform regularization (c.f. Example~\ref{ex:need_diff_reg}), \revisenew{we use the default regularization constant provided in the sklearn package~\cite{gmm_sklearn}. } 
\system adaptive regularization is significantly better than uniform regularization on all datasets.

\vspace{-2mm}
\stitle{Incorporating Transitivity.} We compare incorporating transitivity in a post-processing step with \system's approach of modelling it as constraints in the  learning process. To make postprocessing possible, we need to assume that the left and right table are duplicate free and the matching probability of any tuple pair in the same table to be 0, i.e., $\gamma_{2,3,M} = 0$ in \Cref{eq:transitivity}, and we designate the one tuple pair out of the two cross-table tuple pairs with a higher posterior probability ($\max (\gamma_{1,2,M}, \gamma_{1,3,M})$) as the match. Our way of incorporating transitivity as constraints on posterior probabilities significantly produces better results on all datasets. The difference is particularly huge on the DS dataset, as the left and right tables on the DS dataset are actually not duplicate-free. \system makes no such assumption.

\vspace{-2mm}
\subsection{Sensitivity Analysis}
\label{ssec:sensitivity_analysis}
We show how ZeroER is robust to various hyperparameter settings, features used, and the blocking methods used.

\begin{figure}[htb]
\vspace{-4mm}
  \centering
  \centerline{\epsfig{figure=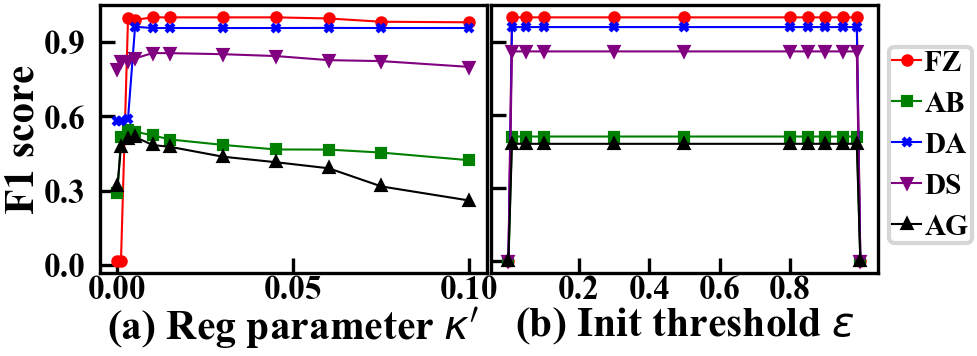, height=3 cm}}
 \vspace{-4mm}
  \caption{F1 score under different (a) regularization parameter $\kappa$, (b) initialization threshold $\epsilon$.}
  \label{fig:reg_init_para}
\vspace{-4mm}
\end{figure}

\vspace{-1mm}
\stitle{Sensitivity to regularization hyperparameter.}
We vary the regularization parameter $\kappa'$ to understand how it affects the results. The Bhattacharyya coefficient is in [0,1], so the regularization parameter has to be in [0,1]. When $\kappa'=1$, the standard deviation of M and U distributions is infinity and the two distribution becomes identical. Therefore, from 0 to $10\%$ of the maximum BC value, i.e. [0,0.1], is a reasonable region for $\kappa'$. 
As shown in \Cref{fig:reg_init_para}(a), the performance of ZeroER is fairly robust for $\kappa'$ in [0,0.1]. Note that $\kappa' = 0$ means there is no feature regularization.
As $\kappa'$ increases, F1 score first increases (regularization helps), then becomes relatively stable for a period,  and finally gradually decreases (too strong of a  regularization). Regardless, $\kappa' = 0.01$ is a good and safe default choice that works well for all datasets.



\vspace{-1mm}
\stitle{Sensitivity to initialization.}
 As shown in \Cref{fig:reg_init_para}(b), \system is robust to initialization threshold $\epsilon$ with no changes on quality for all datasets. When $\epsilon $ = 0 or 1, (which should not be used as initialization parameter value), no data is assigned to M or U class so that EM  fails to recognize two clusters. 

\begin{figure}[htb]
\vspace{-2mm}
  \centering
  \centerline{\epsfig{figure=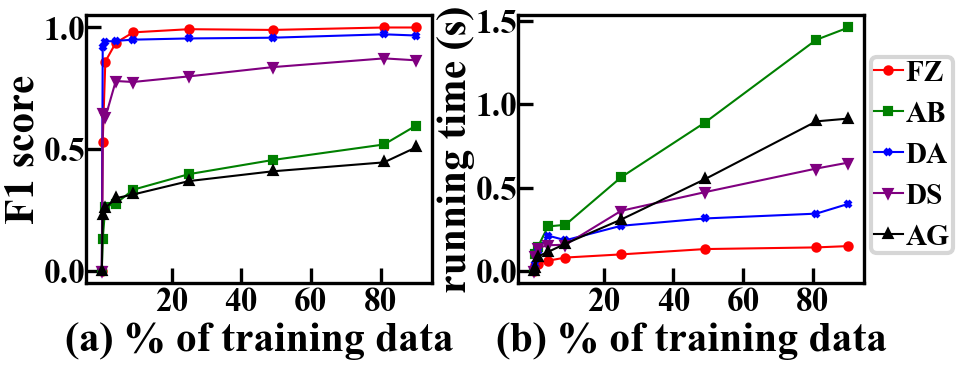,height=3 cm}}
 \vspace{-4mm}
  \caption{(a) F1 score and (b) running time per EM iteration when varying (\%) of unlabeled training data.}
  \label{fig:vary_data}
\vspace{-3mm}
\end{figure}

\vspace{-1mm}
\stitle{Sensitivity \& Scalability to unlabeled training data size.} We demonstrate that ZeroER can produce excellent results
even when trained on a subset of dataset.
Specifically, we vary the amount of data (without labels) used to fit ZeroER and evaluate how it impacts ZeroER's performance on the reminder of data.
As shown in \Cref{fig:vary_data}(a), F1 score increases as the amount of unlabeled training data increases.
\Cref{fig:vary_data}(a) also shows that ZeroER already gives good F1 score with only about 10\% training data, which can save about 90\% training time.
By varying the training data size, we can also demonstrate the scalability of ZeroER. \Cref{fig:vary_data}(b) shows the running time per EM iteration is roughly linear to the amount of training data. This means ZeroER is scalable.


\begin{figure}[htb]
\vspace{-3mm}
  \centering
  \centerline{\epsfig{figure=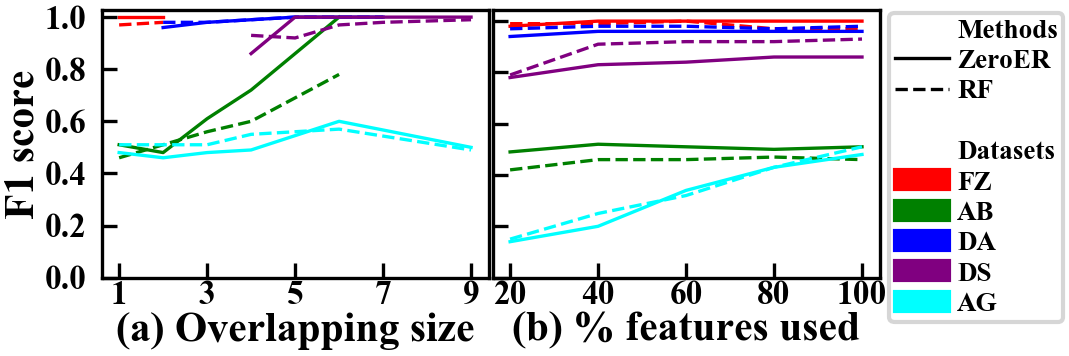,height=2.6cm}}
 \vspace{-4mm}
  \caption{F1 score for \system and RF (random forest) varying (a) degree of blocking and (b) (\%) of features.}
  \label{fig:blocking_feature_sensitive}
\vspace{-3mm}
\end{figure}

\vspace{-1mm}
\stitle{Sensitivity to blocking.}
Though we consider blocking to be orthogonal to us, we vary the aggressiveness of blocking by changing the overlapping size used in the locality sensitive hashing blocking scheme. A more aggressive blocking setting will generate less tuple pairs, but potentially at the risk of missing matches.
We vary the overlapping size from the minimum value (least aggressive) that still allows the algorithm to finish within a preset time limit to the maximum value that results in zero tuple pairs. As shown in \Cref{fig:blocking_feature_sensitive}(a), both ZeroER and RF are robust to blocking. Overall, ZeroER is comparable to RF and is better than random forest on FZ and AB regardless of the overlapping size. \revisenew{By increasing the overlapping size, some of the difficult examples would be excluded in blocking. Dataset AB is affected more dramatically, as it has more difficult examples excluded by blocking.} 

\stitle{Sensitivity to features.}
\revisenew{We vary the amount (\%) of features by sampling from the features generated by Magellan}
to see if the comparison between ZeroER and supervised methods is sensitive to the feature set used. As shown in \Cref{fig:blocking_feature_sensitive}(b) The performance of ZeroER is robust to different (\%) of features on most datasets except AG. Though on AG, both ZeroER and RF have degraded results.

%


\section{Related Work}
\label{sec:relatedWork}

Entity resolution is one of the fundamental and challenging problems in data curation.
A good overview of ER can be found in surveys such as~\cite{DBLP:conf/sigmod/KoudasSS06, DBLP:journals/tkde/ElmagarmidIV07, DBLP:journals/pvldb/DongN09, DBLP:series/synthesis/2010Naumann, DBLP:journals/pvldb/GetoorM12,herzog2007data,maggi2008survey} and tutorials~\cite{DBLP:journals/pvldb/GetoorM12}. Previous work on ER can be broadly divided into three branches depending on how they relate to this paper: (1) identifying useful similarity functions; (2) tackling the efficiency challenges of ER; and (3) designing an algorithm to predict matches/unmatches.

A variety of similarity functions have been used for ER~\cite{DBLP:journals/tkde/ElmagarmidIV07}, including generic ones such as edit distance, jaccard distance, and cosine similarity, as well as some domain specific ones such as Jaro distance for  people names~\cite{JaroDistance}. A comparison of different similarity functions can be found in~\cite{chandel2007benchmarking}. ER work usually either assumes that feature engineering for an ER dataset is done by a human, or employs a default set of similarity functions. \system can work with any similarity functions, and by default, uses features generated by the open-source ER package Magellan~\cite{konda2016magellan}.

A second branch of work aims at reducing the quadratic complexity of an ER task. Some of these algorithms are specific to a similarity function, such as edit distance~\cite{gravano2001approximate} or jaccard distance~\cite{arasu2006efficient,hadjieleftheriou2008fast}. Chaudhuri et al~\cite{chaudhuri2006primitive} provides an efficient similarity join operator that can work with multiple similarity functions. Another main direction, called blocking, aims to avoid comparing tuples pairs that are unlikely to be matches, leveraging techniques such as locality sensitive hashing~\cite{papadakis2016comparative,papadakis2013meta}. \system is orthogonal to this line of research, and takes an optional blocking function as input. 



The third line of related work concerns with designing an algorithm to predict matches. 
Supervised ML approaches provide state-of-the-art results for ER.
However they require substantial amount of
training data in the form of matches and unmatches ~\cite{deeper,anhaisigmod2018,kopcke2010evaluation,dong2018data}. This process is often time consuming, cumbersome and error prone.
Popular approaches include binary classifiers such as Naive Bayes~\cite{Winkler99thestate}, decision tree~\cite{DBLP:conf/vldb/ChaudhuriCGK07} and SVM~\cite{DBLP:conf/kdd/BilenkoM03}. 
Active learning based techniques have also been proposed to reduce the amount of labeled examples~\cite{SarawagiB02,DBLP:dblp_conf/sigmod/ArasuGK10}. We have compared with three representative and best-performing supervised ML models as well as an active learning based approach, and showed that \system achieves comparable results with zero labeled examples. 
General-purpose clustering algorithms (e.g., K-Means and GMM) can be used to perform ER in an unsupervised way. As we have shown, they usually produce much inferior results. 
The Fellegi-Sunter (FS) model~\cite{fellegi1969theory} is generally regarded as the seminal work in probabilistic unsupervised ER, and was famously used for deduplicating the Census data that uses binary features and assumes feature independence~\cite{Winkler99thestate}. Since then, specific adaptations to the FS model has been explored for specific domains, such as US homicides~\cite{sadinle2013generalized},
traffic crashes~\cite{cryer2001investigation}, and hospitalizations~\cite{clark2004practical}. \system can be seen as a generalization of FS model that includes three major innovations (a reduced set of parameters, feature regularization, and incorporating transitivity), which together produces a much more performant unsupervised approach that works across datasets in different domains.

\vspace{-2mm}
\section{Conclusion and Future Work}
\label{sec:conclusion}
In this paper, we proposed \system,
an effective and efficient solution for solving entity resolution with \emph{Zero} labeled examples.
\system provides excellent results across multiple datasets
and is competitive with supervised ML approaches.
This superior performance is achieved by a collection of novel techniques including
generative modeling, adaptive feature regularization and
incorporating transitivity.
\system dramatically reduces the labeling effort of domain scientists in ER tasks.
We hope that \system will play a vital role in the increasingly popular area of self-service data preparation.
\revisenew{Currently, \system treats feature engineering as a black box and uses features generated by Magellan. Designing automatic feature engineering solution for ER in general is an interesting future work.}



\balance
\bibliographystyle{abbrv}
\bibliography{main}  
\end{document}